\documentclass{article}
\usepackage[utf8]{inputenc}

\usepackage[table]{xcolor}
\usepackage[]{tismir}
\usepackage{cite}

\usepackage[super]{nth}

\usepackage[pdftex]{graphicx}
\usepackage{subfig}
\usepackage{pgfplots}
\pgfplotsset{compat=1.18}

\usepackage{float}
\restylefloat{table}

\usepackage{amsmath}
\usepackage{amssymb}
\newcommand{\R}{\mathbb{R}}

\usepackage{array}
\usepackage{booktabs}
\usepackage{multicol}
\usepackage{multirow}
\usepackage{hyperref}
\usepackage{adjustbox}
\usepackage{siunitx}
\usepackage{verbatim}

\usepackage{tablefootnote}

\usepackage{ifthen}
\newcommand{\anonymous}{false}

\usepackage{color}
\newcommand{\gf}{\textcolor{cyan}} 

\newcommand{\change}{}  
\newcommand{\changetable}{}  
\newcommand{\alter}{}  
\newcommand{\remove}[1]{}  
\newcommand{\anonym}{}  
\usepackage[normalem]{ulem}

\title{The Sound Demixing Challenge 2023\\-- Music Demixing Track}

\ifthenelse{\equal{\anonymous}{true}}{}{%
\author{%
Giorgio~Fabbro\thanks{Sony Europe B.V., Stuttgart, Germany}, %
~Stefan~Uhlich\protect\footnotemark[1], %
~Chieh-Hsin~Lai\thanks{Sony Group Corporation, Tokyo, Japan}, %
~Woosung~Choi\protect\footnotemark[2],\\ %
Marco~Martínez-Ramírez\protect\footnotemark[2], %
~Weihsiang~Liao\protect\footnotemark[2], \\ %
Igor~Gadelha\thanks{Moises.ai, João Pessoa, Brazil}, %
~Geraldo~Ramos\thanks{Moises.ai, Salt Lake City, USA}, %
~Eddie~Hsu\protect\footnotemark[3], %
~Hugo~Rodrigues\protect\footnotemark[4], \\%
Fabian-Robert~Stöter\thanks{AudioShake, San Francisco, USA}, %
~Alexandre~Défossez\thanks{Meta AI, Paris, France}, %
~Yi~Luo\thanks{Tencent AI Lab, Shenzhen, China}, %
~Jianwei~Yu\protect\footnotemark[7], \\%
Dipam~Chakraborty\thanks{AIcrowd, Bengaluru, India}, %
~Sharada~Mohanty\thanks{AIcrowd, Lausanne, Switzerland},\\ %
Roman Solovyev\thanks{Institute for Design Problems in Microelectronics of Russian Academy of Sciences, Moscow, Russian Federation}, %
~Alexander Stempkovskiy\protect\footnotemark[10], %
~Tatiana Habruseva\thanks{Independent researcher, Cork, Ireland},\\%
Nabarun~Goswami\thanks{The University of Tokyo,~Japan}\;\;, %
~Tatsuya~Harada\protect\footnotemark[12]$\;\;\;\;\;\;$\thanks{RIKEN,~Japan}\;\;, %
~Minseok~Kim\thanks{Korea University,~Seoul,~South Korea}\;\;, %
~Jun~Hyung~Lee\protect\footnotemark[14],\\ %
Yuanliang~Dong\thanks{Central Conservatory of Music, Beijing, China}\;, 
~Xinran~Zhang\protect\footnotemark[15]\;, 
~Jiafeng~Liu\protect\footnotemark[15]\;\;, 
and~Yuki~Mitsufuji\protect\footnotemark[2]%
}}

\date{}

\type{overview}

\ifthenelse{\equal{\anonymous}{true}}{\authorref{Anonymous,~A.}}{%
\authorref{Fabbro,~G., Uhlich,~S., Lai,~C., Choi,~W., Ramirez,~M., Liao,~W., Gadelha,~I., Ramos,~G., Hsu,~E., Rodrigues,~H., Stöter,~F., Défossez,~A., Luo,~Y., Yu,~J., Chakraborty,~D., Mohanty,~S., Solovyev,~R., Stempkovskiy,~A., Habruseva,~T., Goswami,~N., Harada,~T., Kim,~M., Lee,~J., Dong,~Y., Zhang,~X., Liu,~J., and Mitsufuji,~Y.}}

\ifthenelse{\equal{\anonymous}{true}}{\authorshort{Anonymous, A.}}{\authorshort{Fabbro et al}}
\titleshort{The Sound Demixing Challenge 2023 -- Music Demixing Track}

\begin{document}

\twocolumn[{%
\maketitleblock
\begin{abstract}
This paper summarizes the music demixing (MDX) track of the Sound Demixing Challenge (SDX'23).
We provide a summary of the challenge setup and introduce the task of robust music source separation (MSS), i.e., training MSS models in the presence of errors in the training data.
We propose a formalization of the errors that can occur in the design of a training dataset for MSS systems and introduce two new datasets that simulate such errors: \emph{SDXDB23\_LabelNoise} and \emph{SDXDB23\_Bleeding}\endnote{The datasets are available for download at \url{https://developer.moises.ai/research\#datasets}}.
We describe the methods that achieved the highest scores in the competition.
Moreover, we present a direct comparison with the previous edition of the challenge (the Music Demixing Challenge 2021): the best performing system \alter{\remove{under the standard MSS formulation }}achieved an improvement of over 1.6dB in signal-to-distortion ratio over the winner of the previous competition, when evaluated on MDXDB21.
Besides relying on the signal-to-distortion ratio as objective metric, we also performed a listening test with renowned producers and musicians to study the perceptual quality of the systems and report here the results.
Finally, we provide our insights into the organization of the competition and our prospects for future editions.
\end{abstract}
\begin{keywords}
Music Source Separation, Deep Learning, Neural Networks, Robust Training, Sound, Signal Processing
\end{keywords}
}]
\saythanks{}

\section{Introduction}
\label{sec:intro}
Audio source separation has a long history in research, partially motivated by the many applications it enables.
Thanks to source separation, music producers and artists are able to make use of material that was created decades ago or under limited recording conditions.
Movie production studios have now the possibility to revive old classics\endnote{\url{https://www.youtube.com/watch?v=jcWINJxnw70}}, for which only single-channel master tracks exist, and bring them back to the theatre taking advantage of newer innovations, such as spatial audio~\change{\citep{Petermann2022}}.
\change{Moreover, }people who are experiencing hearing difficulties, a condition which makes it challenging for them to communicate in loud and noisy surroundings, \change{can be supported by this technology and} successfully engage in conversations.

In the context of research, the more recent success of audio source separation should be attributed also to the presence of benchmarks that allowed different methods to be effectively compared: one example is the MUSDB18 dataset~\citep{rafii2017musdb18}, introduced during the Signal Separation Evaluation Campaign (SiSEC) in 2018~\citep{sisec2018}.
The dataset includes 150 different songs, along with corresponding separate recordings for each musical instrument.
These tracks were grouped into four classes for consistency across songs: \emph{vocals}, \emph{bass}, \emph{drums}, and a final class named \emph{other} for any other instrument present.
This allowed participants to carry out training and evaluation of their models on a standardized and consistent pool of data.

In 2021 we continued the tradition of SiSEC competitions with the Music Demixing Challenge (MDX'21)~\citep{mitsufuji2022music}.
We decided to keep MUSDB18 as the reference dataset for training, but introduced a new benchmark for testing based on data that could not be accessed by the participants.
The submissions to the MDX'21 challenge were evaluated on a new hidden dataset called MDXDB21~\citep{mitsufuji2022music}, which contained 30 songs produced by Sony Music Entertainment Japan for the purpose of the challenge.

Two years later, we organized a new edition of the challenge, named Sound Demixing Challenge 2023 (SDX'23): while MDX'21 focused exclusively on music source separation \change{(MSS)}, SDX'23 presented two independent tracks, one for music source separation and one for cinematic sound separation~\citep{uhlich2023TheSoundDemixingCinematicTrack}.
In the music track, we again used MDXDB21 as test benchmark.
This allowed a comparison of the submissions across the two editions of the challenge.

While we wanted to offer the prospective participants a familiar research playground (e.g., keeping the usual four instrument classes introduced by MUSDB18), we complemented that with a novel aspect of the source separation problem: \emph{robust music source separation}.
In robust MSS the trained system needs to be able to handle errors and inconsistencies in the training data.
We decided to focus one part of SDX'23 on this topic after we analyzed the outcome of the previous edition of the challenge.

At the end of MDX'21 it was evident that the volume of data accessible for training can have a significant effect on the quality of the final source separation system.
At the same time, since state-of-the-art methods for source separation are still dominated by supervised learning approaches, the availability of appropriate data is limited.
This formulation requires individual tracks, each containing the signal of a single instrument, which are difficult to obtain, especially in large quantities.
Simulation of the recordings using MIDI and virtual instruments (e.g., the Slakh dataset~\citep{manilow2019cutting}) can be a compromise: the size of the dataset grows very easily, but the results lack realism and have some critical limitations (e.g., they often do not contain vocals).

In the context of the challenge, rather than tackling the problem of the scarcity of the data, we assumed that such data would be available and looked further down the road: is having more songs enough to obtain better models?
How can we best leverage and curate a large corpus of tracks to improve the quality of the separation model?

A set of internal experiments revealed that training source separation models on a large volume of high-quality data does not guarantee better network convergence.
The convergence behavior of our models was dramatically affected by \change{label} errors in the data.
These were mostly related to the ground truth labels used when training (i.e., the identity of the instruments present in each audio recording).
For example, tracks labeled as \emph{vocals} actually contained the signal of a guitar.
Intuitively, the impact of such \change{label} errors should decrease proportionally to the amount of data used to train the model.
Against our expectations, increasing the amount of data was causing the models to stop converging.
Only through an expensive activity of data cleaning we were able to make the model converge again.

\change{While the relative amount of erratic data definitely impacts whether this phenomenon occurs or not, our experiment showed how the difference in performance caused by the errors in the data justifies a closer look into the phenomenon.}
\change{There are no aspects of our experiments that prevent us from generalizing this finding to tasks beyond source separation, or even outside the audio domain.}
\change{Nevertheless, confirming the generalization of this finding would be outside the scope of this work\alter{. W}e leave such activity for future work.}
\change{What can be drawn from our experiments is that an activity that ensures that the data is clean has a visible impact on the final performance of the model.}

The process of cleaning a dataset is very expensive and does not scale easily with its size: manually checking the annotations for a large number of tracks is not feasible.
While automatic methods can provide an initial solution \citep[e.g., audio classification and tagging models:][]{garcia2021leveraging,mahanta2021deep,panns}, some errors are intrinsically difficult to fix: if a single track erroneously contains the signal of two different instruments, simply changing its label will not solve the problem.
Source separation methods can also be used to exclude noisy samples in a large dataset, to ensure that the new model is trained only on clean data~\citep{rouard2023hybrid}.

A more interesting solution would be to make the training of the network invariant to these errors.
Ideally, if the learning process is robust to such inconsistencies, adding new data upon availability becomes easier, as we can avoid a cleaning activity that is expensive, likely incomplete and potentially ineffective.
The research community has proposed some robust training methods to address label noise in classification problems \citep{han2018co,li2020dividemix,cheng2020learning,wang2022promix} or handle out-of-distribution samples \citep{lai2019robust,lai2023robust,hendrycks2021many,mukherjee2021outlier}. However, to the best of our knowledge no existing approach explicitly tackles audio source separation\endnote{One paper \citep{koo2023selfrefining} was published during the preparation of this article which already makes use of our new proposed dataset \emph{SDXDB23\_LabelNoise}. Their approach is not based on training the model on the noisy data, but on automatically correcting the labels in the data before training the model.}.
With the SDX'23 challenge we aimed to bring the attention of the research community to the topic of robust MSS and provided the participants with an environment that presented the issues outlined above, even when working with small datasets.

In this paper we summarize the music track of SDX'23: we show the challenge setup in Sec.~\ref{sec:setup}, we introduce the topic of robust \change{MSS} in Sec.~\ref{sec:robust}, we outline the parts of the challenge related to MSS \alter{without robustness constraints} in Sec.~\ref{sec:unlimited}, we present the challenge results, together with the descriptions of \change{some of} the winning approaches in Sec.~\ref{sec:results} and~\ref{sec:listening_results} and we elaborate on the technical challenges in the organization of the competition in Sec.~\ref{sec:challenge_organization}.

\section{MDX Challenge Setup}
\label{sec:setup}
In the following, we summarize the structure of the competition. Similarly to the previous edition, the challenge was hosted on AIcrowd\endnote{\url{https://www.aicrowd.com/challenges/sound-demixing-challenge-2023}}.

\subsection{Task Definition}
Participants in the music track (MDX) of the Sound Demixing Challenge 2023 were asked to submit systems to separate a stereo song $\mathbf{x}(n) \in \R^2$ into one stereo track for \textit{vocals} ($\mathbf{s}_\text{V}(n) \in \R^2$), one for \textit{bass} ($\mathbf{s}_\text{B}(n) \in \R^2$), one for \textit{drums} ($\mathbf{s}_\text{D}(n) \in \R^2$) and one for \textit{other} ($\mathbf{s}_\text{O}(n) \in \R^2$), where the song can be obtained as:
\begin{equation}
    \mathbf{x}(n) = \mathbf{s}_\text{V}(n) + \mathbf{s}_\text{B}(n) + \mathbf{s}_\text{D}(n) + \mathbf{s}_\text{O}(n),   
\end{equation}
and \change{$n \in \mathcal{Z}$} denotes the time index.
All signals are sampled at 44100Hz.

\subsection{Leaderboards}
The previous edition of the MDX challenge~\citep{mitsufuji2022music} focused on music source separation\alter{, similarly to the SiSEC competitions~\citep{sisec2015,sisec2016,sisec2018}}.
This year, we devoted two of three leaderboards to the issue of training source separation models with data containing errors.
We elaborate on this \change{task} in Section~\ref{sec:robust}.

On top of that, we provided a third leaderboard that was devoted to the \alter{\remove{standard }}music source separation task, without any constraint on the training data.
By allowing training under any condition, we are interested in tracking the progress of the source separation community.
\change{Unlike the previous edition of the challenge, we did not offer any leaderboard for models trained exclusively on MUSDB18~\citep{rafii2017musdb18}.}

In summary, the submissions were categorized under the following three leaderboards:
\begin{itemize}
    \item Leaderboard \change{\emph{LabelNoise}} was designated for models trained on data suffering from \emph{label noise},
    \item Leaderboard \change{\emph{Bleeding}} was designated for models trained on data suffering from \emph{bleeding},
    \item Leaderboard \change{\emph{Standard}} was designated for models trained on any data.
\end{itemize}
For the definitions of \emph{label noise} and \emph{bleeding}, we refer the reader to Sec.~\ref{sec:formalizing_errors}.

\change{It is important to note that the leaderboards \emph{LabelNoise} and \emph{Bleeding} did not allow the usage of any external trained resource, as that would have been equivalent to the usage of external (possibly clean) data.}

\subsection{Ranking Metric}
The evaluation of the systems followed the same strategy as in MDX'21: we used the global \emph{signal-to-distortion ratio} (SDR) as metric, which is defined for one song as
\begin{equation}
    \text{SDR} = \frac{1}{4} \biggl(\text{SDR}_\text{V} + \text{SDR}_\text{B} + \text{SDR}_\text{D} + \text{SDR}_\text{O}\biggr),
    \label{eq:sdr}
\end{equation}
with
\begin{equation}
    \text{SDR}_c = 10\log_{10}\frac{\sum_n\lVert\mathbf{s}_c(n)\rVert^2}{\sum_n\lVert\mathbf{s}_c(n) - \mathbf{\hat s}_c(n)\rVert^2},
\end{equation}
where $\mathbf{s}_c(n) \in \R^2$ and $\mathbf{\hat s}_c(n) \in \R^2$ denote the stereo target and estimate for source $c \in \{\text{V},\text{B},\text{D},\text{O}\}$. Finally, the global SDR of \eqref{eq:sdr} is averaged over all songs in the hidden test dataset to obtain the final score\change{\endnote{Please note that this evaluation metric computes the mean over all time samples and over all songs. This is different from the evaluation metric used in the SiSEC competitions, where the SDR score of a model is the median over the scores on all songs in the test set, and the SDR score on a single song is computed as the median of the SDR scores over segments of one second.}}.

In addition to the objective metric above, this edition of the challenge also featured a subjective evaluation based on a listening test carried out on the estimates of the \alter{top three} systems \change{\alter{in the leaderboard \emph{Standard}}} \alter{(ranked by their SDR score)}.
We refer the reader to Sec.~\ref{sec:listening_test} and Sec.~\ref{sec:listening_results} for details on the subjective evaluation.

\change{The total prize pool for the competition was \$32,000. \$2,000 was reserved for the listening test winner, while the rest was equally distributed among the leaderboards.}

\subsection{Timeline}
The challenge featured two rounds.
Phase~I started on January~\nth{23} 2023, while Phase~II started on March~\nth{6} 2023.
Due to the submission system experiencing difficulties in handling the surge in the number of submissions towards the end of the challenge, the end date of Phase II was extended by one week. Originally scheduled to conclude on May~1st, 2023, the challenge was extended to May 8, 2023, to ensure a fair competition for all teams.
A \emph{warm-up round} was also organized, which began on December~\nth{8} 2022 and lasted until the beginning of Phase~I.
During this round, participants could get acquainted with the submission system and prepare their submissions for the challenge.

\change{To carry out the evaluation of the submissions, we used the same approach as in the previous edition of the challenge.}
\change{The hidden test set (MDXDB21) contains 30 songs, of which we held out three songs: we used two of them to give feedback to the participants about their performance (i.e., they could listen to the separations of their model on them), while we excluded the third one because the bass track is silent.}
\change{The remaining 27 songs were used to compute the scores displayed in the leaderboards.}
\change{During Phase~I, the scores available to the participants were computed on one third of them (nine songs).}
\change{During Phase~II, nine more songs were added (i.e., the scores displayed on the leaderboards were computed over 18 songs).}
\change{These two subsets were selected randomly (we kept the same random subsets used in MDX'21).}
Once the competition ended and the participants could not submit anymore to be eligible for prizes, the final scores computed on the whole test set \change{(27 songs)} were displayed on the leaderboards.

We chose to carry out the evaluation in this way in order to prevent participants from implicitly adapting their submissions to the test set throughout the course of the challenge.
\change{Please note how, at the end of the MDX'21 challenge, all submissions from Phase~II were evaluated on the 27 songs to produce the scores on the final leaderboards.}
\change{T}his encouraged participants to maximize the number of submissions during Phase~II, so that they would increase the likelihood \change{of one of them achieving a high score on the full test set}.
For this reason, at the end of Phase~II \change{of SDX'23,} we asked each participant in every leaderboard to \emph{manually} select three candidate submissions that would move on to the final evaluation: only these submissions were then evaluated on the whole test set. The best out of three was then displayed as the final entry for each team.





\section{Robust Music Separation}
\label{sec:robust}
From the results of the last edition of the challenge \citep{mitsufuji2022music} it became clear to the source separation community that the performance of a model very often correlates with the amount of data used to train it.
One example was the participant \emph{defossez}: his model\endnote{\url{https://github.com/facebookresearch/demucs/tree/v3}}~\citep{rouard2023hybrid} trained only on MUSDB18 achieved an average SDR score of 7.32dB\endnote{\url{https://www.aicrowd.com/challenges/music-demixing-challenge-ismir-2021/leaderboards?challenge_leaderboard_extra_id=868&challenge_round_id=886}}, while training on additional data improved its performance by almost 1dB\endnote{\url{https://www.aicrowd.com/challenges/music-demixing-challenge-ismir-2021/leaderboards?challenge_round_id=886}}.

It would be safe to assume that increasing the amount of data used for training increases the performance of the model, but this is not necessarily the case.
While performing experiments using internal data, we experienced that \change{a model} whose validation loss used to converge when trained on a small high-quality \change{(i.e., clean)} dataset \change{was} not converging anymore when trained on a high-quality dataset one order of magnitude larger (see Figure~\ref{fig:dataset_comparison}).
After careful inspection, we realized that \change{label} errors in \change{this} large pool of data were responsible for this phenomenon.
\change{Please note that the general practice of increasing the amount of training data to increase performance still applies. Once the large pool of data had been cleaned, we achieved a higher performance than with the smaller dataset.}

\subsection{Why Robust Music Separation?}
\begin{figure}[ht]
    \centering
    \includegraphics[width=0.5\textwidth]{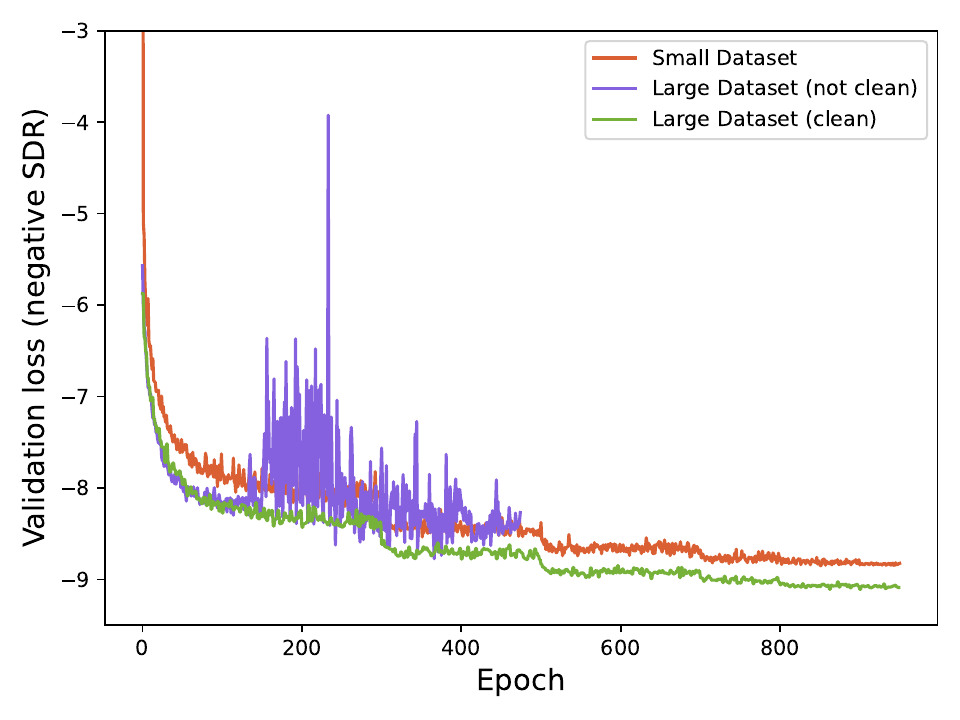}
    \caption{Comparison of validation loss when training the same model on a small dataset (red), a large dataset with errors (purple) and the same large dataset once the errors have been corrected (green). All experiments were evaluated on the same validation set.}
    \label{fig:dataset_comparison}
\end{figure}

Intuitively, the more data we have, the lower the impact of any incorrect recording should be.
However, increasing the amount of data also increases the number of incorrect recordings and the likelihood that globally the dataset presents inconsistencies.
Examples of such errors can be recordings where the recorded instrument does not match the instrument label, or individual recordings that contain the signal of more than one instrument.
As Figure~\ref{fig:dataset_comparison} shows, only through an expensive activity of \emph{data cleaning} we were able to make those models converge again, confirming that those \change{label} errors were the cause of the issue.

Given how time-consuming and resource-expensive a data cleaning activity is, we decided to focus the research community on this issue during the Sound Demixing Challenge 2023.
Our intention was to constrain the participants in the data at their disposal, so as to make them devise strategies that would allow training models \emph{robust} to inconsistencies in the training data.

\subsection{Formalizing Errors in the Training Data}
\label{sec:formalizing_errors}
We aimed at providing a common framework to characterize errors and inconsistencies in the training data from the perspective of audio source separation.
We assumed that the training of a source separation model requires individual stems and does not rely exclusively on mixtures.
Under these assumption, the integrity of the individual recordings becomes paramount.
We identified two categories of errors that can occur:
\begin{itemize}
    \item the recorded instrument and the label identifying it do not agree,
    \item the recording contains signals belonging to more than one instrument.
\end{itemize}

We named the first category \textbf{label noise}, since a (possibly) random change in the label of the recording is responsible for it; we named the second category \textbf{bleeding}, as it occurs when the signal of a second instrument \textit{bleeds} into the recording of the first one.
\change{Two leaderboards} in the challenge were dedicated to \change{robust MSS, one for each of} the two error categories \change{(\emph{LabelNoise} and \emph{Bleeding})}.

\paragraph*{Label Noise}
Being a creative process, music production does not follow conventional workflows.
Different music producers adhere to different conventions for their procedures, which has direct consequences in how their final deliverable is organized.
Often producers are asked to deliver not only the final mix of their song, but also individual stems: these tracks usually group multiple recordings that belong to a single identifiable instrument (e.g., all microphones for the drum set).
The choice of which stems to deliver and how to name them is typically influenced by the preferences and background of the producer and by the style of the song.
In the context of music production there is no immediate benefit in following a strict convention when naming the stems.

When we repurpose these recordings for training source separation models, we need to collect stems coming from different music producers.
For one stem, the only information we require is the identity of the recorded instrument.
Typically, the only label we possess that indicates the instrument identity in a recording is its file name.
Some producers follow more organized workflows and might keep some extra metadata, but this does not hold for all of them.
The stem file name is chosen manually: a process that is prone to errors and not conforming to any naming convention.
The result is a very large collection of valuable audio recordings and a chaotic set of instrument names.

Even if we were able to efficiently collapse all the variations of one instrument name (e.g., \emph{el\_guitar}, \emph{electric\_guitar}, \emph{el\_gtr}, etc.), we would still have some which are intrinsically ambiguous.
Some examples are: \emph{lead} (which could refer to vocals, guitar, or any other instrument playing a leading part), \emph{choir} (which could be a human choir or a synthesizer imitating a choir), \emph{bells} (which could refer to church bells, chime bells or a synthesizer sound), attributes used as names (such as \emph{clean}, \emph{dark}, \emph{bright}), \emph{sfx} (which could potentially refer to any sound), and others.


\paragraph*{Bleeding}
When producing music in a studio, the priority is given to the performance.
Each performer must be in the condition of delivering the perfect rendition of the artist's vision.
One example is for the whole band to play together, instead of recording one instrument at a time.
This improves the musical interpretation of the song, as each musician can directly react to small changes in the performance of the others: this leads to increased interactions that translate into a more truthful and lively recording.
Recording studios are designed to maximize acoustic isolation between different rooms and booths, but there are limits.
For example, low frequencies are notoriously difficult to isolate, due to their long wavelength: a bass amplifier produces sound that can overcome the isolation barriers in the studio and reach the microphones devoted to other instruments.
If those instruments are recorded at the same time, some signal from the bass amplifier will \emph{bleed} into their tracks.
\change{This situation arises also when recording performances with an orchestra~\citep{orchestrableeding}.}
This is not an issue in the context of the song production, as those signals will be summed anyway to produce the final mix.
But if these recordings are now repurposed as training material for a source separation system, this bleeding becomes problematic.


\subsection{Creating Datasets with Corruptions}
In the context of the challenge, we wanted participants to tackle the issues above while ensuring that the overall competition remained fair when systems are evaluated against a common test set.
Ideally, we would force the participants to train their systems using the same datasets, which have been corrupted using label noise and bleeding.
On top of that, we would need to prevent access to an error-free version of those datasets.

For both categories of errors we have the option of simulating them on existing recordings.
Although simulation might result in a loss of realism, it preserves the conditions required during the setup of the challenge and does not change the underlying task that needs to be solved.
For this reason, we decided to simulate both label noise and bleeding starting from error-free recordings.

Although clean recordings are easy to find in the community, we needed to avoid participants having access to the source material, as it would have given them an unfair advantage.
For this reason, we made use of MoisesDB~\citep{pereira2023moisesdb}, which was not released until after the end of the challenge.
It contains 240 individual tracks sourced from 45 diverse artists, spanning twelve distinctive musical genres. Each track is broken down into its constituent audio components and is categorized within a hierarchical, two-tier taxonomy of stems\change{~\citep{Manilow2020HierarchicalMI}}.

We selected 203 songs from MoisesDB to be the source data for our corrupted datasets.
Then, we generated two versions of such data, one containing label noise (\emph{SDXDB23\_LabelNoise}), another containing bleeding (\emph{SDXDB23\_Bleeding}), and shared them with the participants.
The new datasets follow the same structure and design of MUSDB18 and MDXDB21, where each song is composed of four stems: \emph{vocals}, \emph{bass}, \emph{drums} and \emph{other}.
The original song can be recovered as the summation of the four stems\endnote{Please note that summing the four stems will not yield the exact mixture for \emph{SDXDB23\_Bleeding} as we simulate it by adding bleeding components to the original stems.}.
We make both datasets available for download\endnote{\url{https://developer.moises.ai/research\#datasets}}.

When simulating the errors, our objective was to cause a degradation in the training loss of Open-Unmix~\citep{stoter2019open} (upon convergence) of approximately 1dB SDR, when trained individually on label noise and bleeding, compared to training the same model on error-free data.
Moreover, we chose not to split the datasets into a training and validation part \change{to give the participants more freedom to decide how (and whether) to perform validation}.

\paragraph*{Label Noise}
\begin{figure}[ht]
    \centering
    \includegraphics[width=0.5\textwidth]{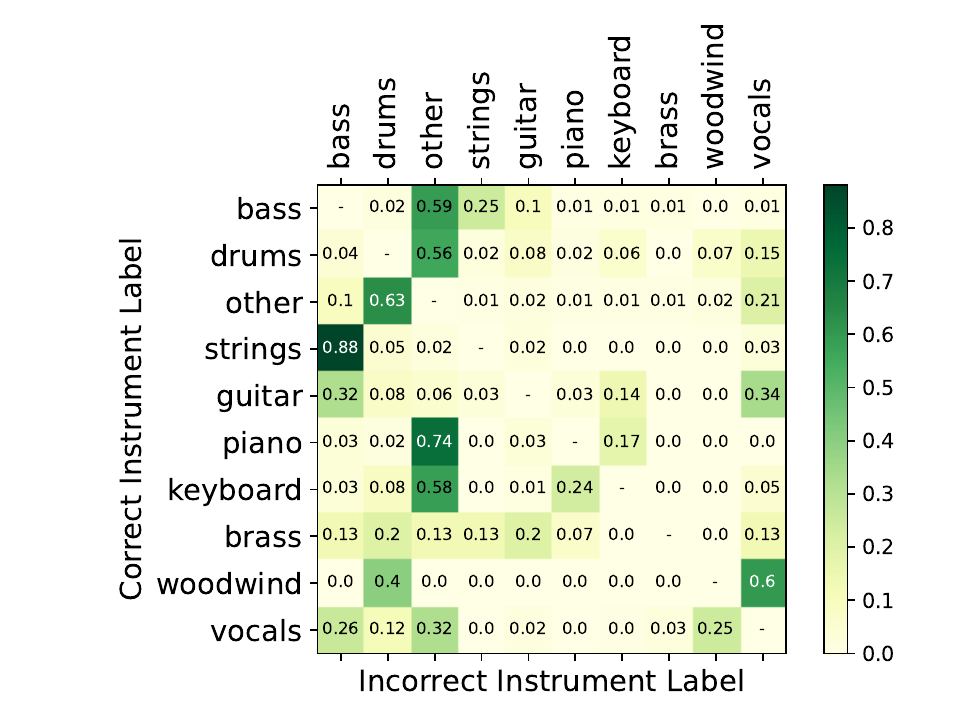}
    \caption{Statistics collected during our internal data cleaning activity. The values in the rows are normalized so that they sum to 1. For example, in all the errors we found in our internal data, the chance that a guitar was labeled as bass is \SI{32}{\percent}.}
    \label{fig:instr_label_errors}
\end{figure}

The simulation of label noise is based on randomly changing the label of a stem.
This is applied on a subset of the stems in the clean dataset: this allowed us to effectively simulate label noise as occurs in real life.
We chose to apply label noise to \SI{20}{\percent} of the stems in the source data, independently of their label.
If a stem was subject to label noise, the new label was sampled from a distribution that reflected the label noise we found in our internal datasets.
To do this, we collected statistics over the frequency of corrections we performed during our data cleaning activity: if a stem initially labeled as $c_\mathit{src}$ was corrected to $c_\mathit{dst}$, we increased the likelihood that during the simulation of label noise we would change $c_\mathit{dst}$ into $c_\mathit{src}$.
These statistics are reported in Figure~\ref{fig:instr_label_errors}.

Please note that our statistics refer to a taxonomy of ten musical instruments, while the final dataset contains stems for four: this means that some corruptions we perform will not have an effect in the final version of the dataset.
For example, changing \emph{drums} into \emph{bass} represents an error in the final dataset, changing \emph{guitar} to \emph{piano} does not (as they both belong to the class \emph{other}).
In the end, \SI{34}{\percent} of the stems in \emph{SDXDB23\_LabelNoise} were affected by label noise\footnote{Please note how, despite we applied label noise only on \SI{20}{\percent} of the stems in the source data, since several stems were grouped into one (\textit{other}) for every song, the relative amount of corrupted stems increased to \SI{34}{\percent}.}.

\paragraph*{Bleeding}
Simulating bleeding was a more elaborate process. 
The amount of bleeding in a recording is usually low (in terms of signal-to-noise ratio).
In order to achieve the target impact of 1dB SDR on the model convergence, we decided to apply corruptions to \emph{every single stem} in \emph{SDXDB23\_Bleeding}.
More specifically, \textit{every stem bleeds into every other stem} in the same song.

The bleeding component in a stem was obtained as a copy of another stem, where we applied gain reduction and filtering.
The scaled and filtered signal was then summed to the stem that would contain the corruption.
We randomly scaled each bleeding signal in the range $[-7, -12]$dB and applied either a low-pass filter or a band-pass filter (we randomly selected for every stem).
The order of the filter was randomly chosen in the range $[3, 10)$.
When we applied a low-pass filter, the cut-off frequency was randomly chosen in the range $[900, 9000)$Hz.
When we applied a band-pass filter, we randomly chose the low and high cut-off frequencies in the range $[200, 600]$Hz and $[8, 10]$kHz, respectively.
All random choices were drawn from uniform distributions.
The ranges were designed empirically to compromise between bleeding realism and desired impact on the model convergence.

\subsection{Robust Baseline Model}
We present here a simple baseline model for the task of robust MSS, which can be used for both label noise and bleeding.
This method is invariant to the choice of network architecture: in our experiments we used Open-Unmix~\citep{stoter2019open}.

We first trained the model on the full noisy dataset $D_1$, without any data cleaning, and obtained a system that achieved suboptimal performance: we name this $\text{\emph{UMX}}^{(1)}$.
We then ran inference of $\text{\emph{UMX}}^{(1)}$ \textit{on the same data it was trained upon}.
In other words, we used $\text{\emph{UMX}}^{(1)}$ to remove some of the errors present in $D_1$: although the model had only suboptimal performance, we could expect it to be able to remove part of the errors in the data.
This step created a new improved version of $D_1$, which we name $D_2$.
We then trained a new model $\text{\emph{UMX}}^{(2)}$, from scratch, on $D_2$: we expected this model to achieve better performance than $\text{\emph{UMX}}^{(1)}$. 
We could then repeat this iterative refinement of the training data $N$ times and train the final model $\text{\emph{UMX}}^{(N)}$.
The maximum number of iterations $N$ was found empirically: to realize our baseline, we used $N = 2$.
Please note how this method can be interpreted as a distillation approach, where the current model $\text{\emph{UMX}}^{(i)}$ acts as a teacher during the training of the next model $\text{\emph{UMX}}^{(i+1)}$ (the student)\change{~\citep{hinton2015distilling,Wang2021SemiSupervisedSV,luo2023music}}.

\begin{figure*}
    \centering
    \subfloat[\alter{The \emph{filtered} method: each stem in the cleaned dataset (i.e.,~at iteration $i+1$) contains only contributions of the corresponding stem in the noisy dataset (i.e.,~at iteration $i$).}]{\includegraphics[width=0.45\textwidth]{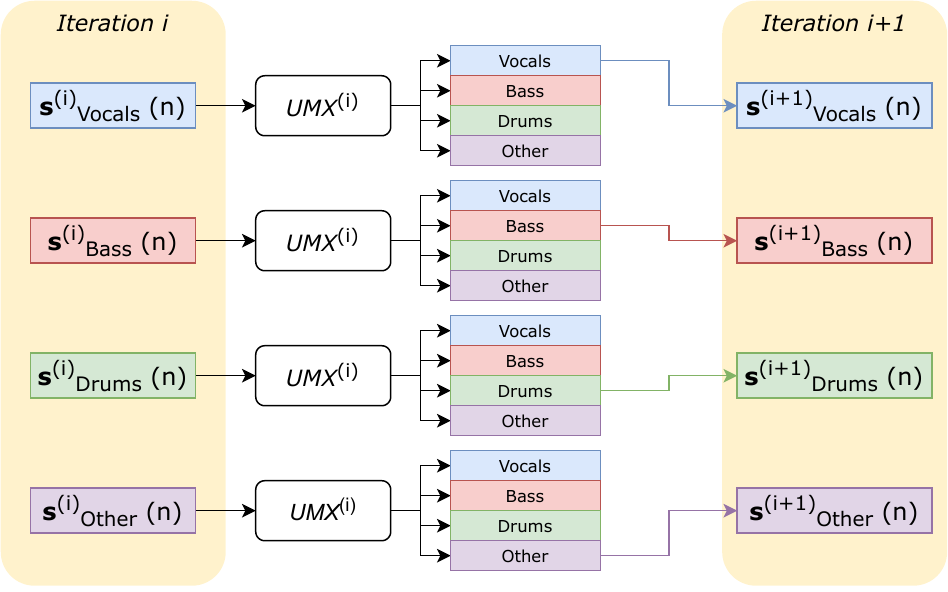}\label{fig:robust_baseline_filtered}}
    \hfill
    \subfloat[\alter{The \emph{redistributed} method: each stem in the cleaned dataset (i.e.,~at iteration $i+1$) is the sum of contributions coming from all the stems of the corresponding song in the noisy dataset (i.e.,~at iteration $i$).}]{\includegraphics[width=0.45\textwidth]{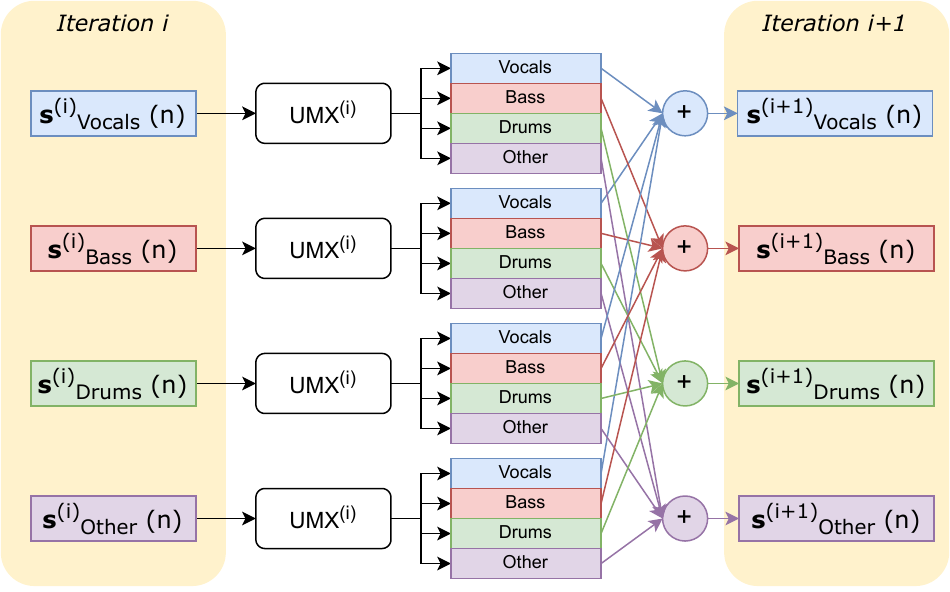}\label{fig:robust_baseline_redistributed}}
    \caption{\alter{The process of cleaning the stems of one song in the noisy dataset using the proposed robust baseline model. We propose two different methods: \emph{filtered} and \emph{redistributed}.}}
    \label{fig:robust_baseline}
\end{figure*}

The creation of the improved dataset at every iteration can be performed in two ways: we name them \alter{\emph{filtered} and \emph{redistributed}}.
Every song in the dataset $D_i$ at iteration \alter{$0 < i \leq N$} is composed of stems \alter{$\mathbf{s}^{(i)}_{c}(n), \forall c \in C$, where $c$} is the \alter{label of the stem} \change{and $C = \{\text{V},\text{B},\text{D},\text{O}\}$ is the set of all sources}.
These stems are used as input to the current model $\text{\emph{UMX}}^{(i)}$ to create $D_{i+1}$.
\alter{We denote the separation of instrument $c \in C$ with the current model as $\text{\emph{UMX}}^{(i)}_c (\cdot)$.}

In the \emph{filtered} method we define a new stem \alter{with label $c$} as the following:
\begin{equation}
    \alter{\mathbf{s}^{(i+1)}_{c}(n) = \text{\emph{UMX}}^{(i)}_c ( \mathbf{s}^{(i)}_{c}(n) ),}
\end{equation}
that is, we only consider as input the stem of the current instrument $c$.
Any contribution of the current instrument $c$ present in the other stems of the same song is discarded \alter{(i.e.,~we discard all contributions of instrument $c$ present in $\mathbf{s}^{(i)}_{\bar{c}}(n), \forall \bar{c} \neq c$)}.

In the \emph{redistributed} method we define a new stem \alter{with label $c$} as the following:
\begin{equation}
    \alter{\mathbf{s}^{(i+1)}_{c}(n) = \sum_{\bar{c} \in C} \text{\emph{UMX}}^{(i)}_c ( \mathbf{s}^{(i)}_{\bar{c}}(n) ),}
\end{equation}
that is, we sum together all the estimates for one instrument $c$ given as input the stems in the current song \alter{for \emph{every} instrument $\bar{c} \in C$}
.
The contributions of the current instrument $c$ present in the wrong input stems (i.e., when $\bar{c} \neq c$) are preserved in the \alter{new dataset and placed in the correct stem}. 
\alter{We provide a visual comparison of the \emph{filtered} and \emph{redistributed} methods in Figure~\ref{fig:robust_baseline}.}

\change{Please note that the model described above was not provided to the participants as a baseline during the challenge.}
\change{To supply the participants with reference scores during the competition, we trained models separately on \emph{SDXDB23\_LabelNoise} and \emph{SDXDB23\_Bleeding} without using any robust method: we provided two baselines based on Open-Unmix~\citep{stoter2019open}, two based on Hybrid Demucs~\citep{hdemucs} and two based on MDX-Net~\citep{mdxnet}.}


\section{Standard Music Separation}
\label{sec:unlimited}

Part of the success of the last edition of the challenge was due to the fact that not only researchers in academia but also industry players participated and submitted their systems.
This was made possible by the presence of a leaderboard where no constraint was given: any model of any complexity, trained on any data, could participate\endnote{We only imposed a limit on the model's inference time on a GPU.}.

For this edition, we offered the same in \change{the leaderboard \emph{Standard}}.
Our intention was to measure the improvement in the performance of the submitted systems with respect to the same benchmark in 2021.

\subsection{Baseline Models}
We provided several baselines to the participants \change{also for the \alter{leaderboard \emph{Standard}}}.
The role of a baseline model was not only to appear in the leaderboard as a reference for existing methods; each participant could choose also to use it as starting point for their submissions.

We provided Open-Unmix~\citep{stoter2019open} (the large model, \emph{\mbox{UMX-L}}), Band-Split RNN \change{(BSRNN)}~\citep{luo2023music} and \change{Cross-Net-Open-Unmix}~\citep{sawata2021all} (\emph{\mbox{X-UMX-M}}).
The former was already provided for the previous edition of the challenge, while the other two were new.

As described in the original paper~\citep{luo2023music}, to create the BSRNN baseline we trained 4 separate BSRNN models for the four instruments in the MUSDB18 dataset~\citep{rafii2017musdb18}.
We used the same band-split schemes for the four tracks, but used slightly smaller model sizes.
More specifically, we used 10 BSRNN blocks instead of 12, and set the feature dimension to 80 instead of 128.
The full pipeline as well as the model weights are available online\endnote{\url{https://gitlab.aicrowd.com/Tomasyu/sdx-2023-music-demixing-track-starter-kit}}.

The \emph{\mbox{X-UMX\change{-M}}} model \change{is an extension of Open-Unmix~\citep{stoter2019open} where the activations of the four sources are averaged before and after the recurrent layers.
This enables the exchange of information across the activations for different sources (hence ``Cross-Net'').
The baseline we provide} corresponds to the one trained on 20k songs \change{(hence the ``M'' in the name)}, as described by~\citet{sawata2023whole}.

\subsection{Listening Test}
\label{sec:listening_test}
For non-robust MSS only, in addition to the objective evaluation based on SDR, we organized a subjective evaluation based on a listening test where professionals in the music industry rated the separations of the best systems.
The main goal of the listening test was to assess the performance of various source separation models via comparative AB sampling.
We included in the listening test one model for each of the three teams that achieved the highest SDR scores on \change{the \emph{Standard}} leaderboard (\emph{SAMI-ByteDance}, \emph{ZFTurbo} and \emph{kimberley\_jensen}).
During the listening test, each assessor was in charge of judging segments of either the output of a source separation model or its residual (i.e., we subtract the output of the model from the input mixture).
The latter reveals how well the model can suppress a source, highlighting potential residues of the signal of interest that are not correctly suppressed.

The listening test was carried out \change{online, using a web interface developed specifically for this occasion,} by a panel of seven assessors who bring experience from various domains of music.
This panel comprised award-winning singers, songwriters, composers, music producers, sound engineers, and an educator.
The assessors were adequately trained to understand and identify common issues associated with source separation, such as distortion and artifacts.
\change{The online interface featured a brief explanation of the purpose of the test, a description of the interface and recommendations to use high-grade audio equipment\endnote{Please note that, since the test was run online, we did not have control over the equipment that the assessor used to perform the test.}.}
Each assessor was expected to complete a minimum of 72 comparisons \change{in total (possibly in multiple sessions)}, equivalent to three comparisons for each pair of models, across the four instrument classes plus their residuals.

To generate the evaluation data for the test, we selected ten songs from those in MoisesDB not used to create \emph{SDXDB23\_LabelNoise} and \emph{SDXDB23\_Bleeding}.
\change{We intentionally selected songs belonging to six different genres, to maximize their diversity.}
We applied the three candidate models to each song and obtained the separated signals.
\change{In order to keep the duration of the listening test feasible, we} did not have the assessors evaluate the complete separated \change{songs}.
Rather, we identified in each song four segments of three musical bars each, where the audio energy was sufficiently high, and we used only the separated signals of those segments.
We then created the residual signals of the separations.
Finally, we paired separations of the same segment obtained with different models and randomly assigned them to the assessors.
\section{Challenge Results and Winning Approaches}
\label{sec:results}
In this section we provide the results of the MDX track of the challenge. 
Tables~\ref{tab:ldbA},~\ref{tab:ldbB} and~\ref{tab:ldbC} show the final scores on the challenge leaderboards.
\change{For each leaderboard, we also report which participants received a prize.
This does not correspond to the final ranking as, in order to receive a prize, the participant needed to disclose their solution.}
\alter{We report in section~\ref{sec:iterative_baseline_results} the results of our iterative refinement baseline for robust MSS.}

\alter{After the conclusion of the challenge, we contacted the top four teams in each leaderboard and invited them to contribute to this manuscript with a description of their approaches.
Sections~\ref{sec:team_zfturbo},~\ref{sec:team_subatomicseer},~\ref{sec:team_CCOM} and~\ref{sec:team_kuielab} report the systems of those teams who accepted our invitation and are written by the corresponding teams.}
\alter{Please note that, despite the selected teams participating in all leaderboards, we choose to describe below only their most significant contributions.
For example, when one team proposed an interesting approach for robust MSS, we mainly describe their solutions for leaderboards \emph{LabelNoise} and \emph{Bleeding}, while for teams that achieved very high scores on non-robust MSS, we report their solution for leaderboard \emph{Standard} and neglect their solutions to the other two leaderboards.
This way, we aim to provide more insight into the scientific choices that allowed the team to achieve higher performance.
At the beginning of each section we therefore indicate what kind of contribution is reported.
}

\begin{table*}
\centering
\renewcommand{\arraystretch}{1.2}
\begin{tabular}{@{}rlcccccclcc@{}}
\toprule
\multirow{2}{*}{\textbf{Rank}} & \multirow{2}{*}{\textbf{Participant}} & \multirow{2}{*}{\textbf{Prize}} & \multicolumn{5}{c}{\textbf{Global SDR} (dB)} & \phantom{abc} & \multicolumn{2}{c}{\textbf{Submissions to \change{\emph{LabelNoise}}}} \\
\cline{4-8}\cline{10-11}
& & & Mean & Bass & Drums & Other & Vocals & & 1st phase & 2nd phase \\
\midrule
\multicolumn{2}{@{}l@{}}{\emph{Submissions}}\\
1. & CCOM & \nth{1} & \textbf{7.46} & \textbf{8.12} & \textbf{7.99} & \textbf{5.34} & \textbf{8.37} & & 7 & 50 \\
2. & subatomicseer & \nth{2} & 6.60 & 6.67 & 7.03 & 4.61 & 8.07 & & 65 & 33 \\
3. & kuielab & \nth{3} & 6.51 & 6.71 & 6.71 & 4.82 & 7.82 & & 99 & 25 \\
4. & aim-less & & 6.44 & 6.75 & 7.19 & 4.56 & 7.28 & & 10 & 22 \\
5. & yang\_tong & & 6.33 & 6.29 & 7.46 & 3.94 & 7.65 & & - & 2 \\[0.1cm]
\multicolumn{2}{@{}l@{}}{\emph{Baselines}}\\
 & UMX & & 3.01 & 3.77 & 2.84 & 1.62 & 3.83 \\
 & Demucs & & 4.84 & 5.55 & 5.68 & 2.89 & 5.23 \\
 & MDX-Net & & 3.49 & 4.26 & 2.84 & 2.42 & 4.42 \\
\bottomrule
\end{tabular}
\caption{Final \change{\emph{LabelNoise}} leaderboard (models trained only on \emph{SDXDB23\_LabelNoise}; top-5)}
\label{tab:ldbA}
\end{table*}

\begin{table*}
\centering
\renewcommand{\arraystretch}{1.2}
\begin{tabular}{@{}rlcccccclcc@{}}
\toprule
\multirow{2}{*}{\textbf{Rank}} & \multirow{2}{*}{\textbf{Participant}} & \multirow{2}{*}{\textbf{Prize}} & \multicolumn{5}{c}{\textbf{Global SDR} (dB)} & \phantom{abc} & \multicolumn{2}{c}{\textbf{Submissions to \change{\emph{Bleeding}}}} \\
\cline{4-8}\cline{10-11}
& & & Mean & Bass & Drums & Other & Vocals & & 1st phase & 2nd phase \\
\midrule
\multicolumn{2}{@{}l@{}}{\emph{Submissions}}\\
1. & kuielab & \nth{1} & \textbf{6.58} & \textbf{6.98} & 6.65 & \textbf{4.96} & \textbf{7.74} & & 99 & 13 \\
2. & ZFTurbo & \nth{2} & 6.38 & 6.94 & \textbf{6.86} & 4.62 & 7.12 & & 32 & 4 \\
3. & subatomicseer & \nth{3} & 6.31 & 6.33 & \textbf{6.86} & 4.59 & 7.47 & & 65 & 11 \\
4. & CCOM &  & 6.20 & 6.34 & 6.32 & 4.28 & 7.87 & & 7 & 17 \\
5. & alina\_porechina &  & 5.87 & 6.01 & 6.10 & 4.09 & 7.30 & & 99 & 118 \\[0.1cm]
\multicolumn{2}{@{}l@{}}{\emph{Baselines}}\\
 & UMX &  & 3.61 & 3.90 & 3.85 & 2.50 & 4.17 \\
 & Demucs &  & 5.33 & 5.90 & 5.56 & 3.69 & 6.19 \\
 & MDX-Net &  & 3.56 & 4.00 & 2.30 & 2.65 & 5.29 \\
\bottomrule
\end{tabular}
\caption{Final \change{\emph{Bleeding}} leaderboard (models trained only on \emph{SDXDB23\_Bleeding}; top-5)}
\label{tab:ldbB}
\end{table*}

\begin{table*}
\centering
\renewcommand{\arraystretch}{1.2}
\begin{tabular}{@{}rlcccccclcc@{}}
\toprule
\multirow{2}{*}{\textbf{Rank}} & \multirow{2}{*}{\textbf{Participant}} & \multirow{2}{*}{\textbf{Prize}} & \multicolumn{5}{c}{\textbf{Global SDR} (dB)} & \phantom{abc} & \multicolumn{2}{c}{\textbf{Submissions to \change{\emph{Standard}}}} \\
\cline{4-8}\cline{10-11}
& & & Mean & Bass & Drums & Other & Vocals & & 1st phase & 2nd phase \\
\midrule
\multicolumn{2}{@{}l@{}}{\emph{Submissions}}\\
1. & SAMI-ByteDance &  & \textbf{9.97} & \textbf{11.15} & \textbf{10.27} & \textbf{7.08} & \textbf{11.36} & & 13 & 5 \\
2. & ZFTurbo & \nth{1} & 9.26 & 9.94 & 9.53 & 7.05 & 10.51 & & 32 & 24 \\
3. & kimberley\_jensen & \nth{2} & 9.18 & 10.06 & 9.47 & 6.80 & 10.40 & & 86 & 134 \\
4. & kuielab & \nth{3} & 8.97 & 9.72 & 9.43 & 6.72 & 10.01 & & 99 & 54 \\
5. & alina\_porechina &  & 8.63 & 9.92 & 9.29 & 6.23 & 9.07 & & 99 & 172 \\[0.1cm]
\multicolumn{2}{@{}l@{}}{\emph{Baselines}}\\
 & UMX-L &  & 6.52 & 6.62 & 6.84 & 4.89 & 7.73 \\
 & BSRNN &  & 6.14 & 5.63 & 6.53 & 4.43 & 7.98 \\
 & X-UMX-M &  & 6.30 & 5.85 & 6.87 & 4.42 & 8.04 \\
\bottomrule
\end{tabular}
\caption{Final \change{\emph{Standard}} leaderboard (models trained on any data; top-5)}
\label{tab:ldbC}
\end{table*}

\subsection{Iterative Refinement Baseline}
\label{sec:iterative_baseline_results}

\begin{table*}[ht]
\centering
\renewcommand{\arraystretch}{1.2}
\begin{tabular}{@{}rlccccclcc@{}}
\toprule
 & & \multicolumn{5}{c}{\textbf{Global SDR} (dB)} \\
\cline{3-7}
& & Mean & Bass & Drums & Other & Vocals \\
\midrule
\multicolumn{2}{@{}l@{}}{MoisesDB (203 songs)}\\
 & Original dataset & 4.43 & 4.65 & 5.06 & \textbf{3.02} & \textbf{5.00} \\
 & Improved dataset (redistributed) & 4.27 & 4.68 & 4.93 & 2.72 & 4.75 \\
 & Improved dataset (filtered) & \textbf{4.46} & \textbf{5.07} & \textbf{5.16} & 2.77 & 4.86 \\
\midrule
\multicolumn{2}{@{}l@{}}{\emph{SDXDB23\_LabelNoise}}\\
 & Original dataset & 3.01 & 3.76 & 2.83 & 1.62 & 3.82 \\
 & Improved dataset (redistributed) & 3.44 & 4.00 & 3.81 & 1.86 & 4.08 \\
 & Improved dataset (filtered) & \textbf{3.90} & \textbf{4.57} & \textbf{4.57} & \textbf{2.22} & \textbf{4.25} \\
\midrule
\multicolumn{2}{@{}l@{}}{\emph{SDXDB23\_Bleeding}}\\
 & Original dataset & 3.60 & 3.90 & 3.84 & 2.50 & 4.17  \\
 & Improved dataset (redistributed) & 3.59 & 3.73 & 4.07 & 2.40 & 4.17  \\
 & Improved dataset (filtered) & \textbf{4.09} & \textbf{4.65} & \textbf{4.76} & \textbf{2.52} & \textbf{4.44} \\[0.1cm]
\bottomrule
\end{tabular}
\caption{Results of our iterative refinement baseline. We use a source separation algorithm trained on corrupted data to improve the dataset: training the same model on the improved data increases the separation quality.}
\label{tab:iterative_baseline_results}
\end{table*}


We report in Table~\ref{tab:iterative_baseline_results} the performance of our iterative refinement baseline.
First of all, we highlight the impact that the errors in the data have on the performance of the model: training on \emph{SDXDB23\_LabelNoise} degrades the average separation quality by \change{1.42}dB, while training on \emph{SDXDB23\_Bleeding} degrades it by \change{0.83}dB.

We then use these models to improve the training data, first by using our \emph{redistributed} approach.
If we train a new model on this improved dataset, average performance on \emph{SDXDB23\_LabelNoise} increases by \change{0.43}dB in SDR, while we experience virtually no change for \emph{SDXDB23\_Bleeding}.

If we use our \emph{filtered} approach to improve the training data, we see an average improvement of \change{0.89}dB for \emph{SDXDB23\_LabelNoise} and of \change{0.50}dB for \emph{SDXDB23\_Bleeding}.

Please note that we also report scores for when we apply our iterative approach to the clean data only (\emph{MoisesDB}): we experience a loss of \change{0.16}dB only when using the \emph{redistributed} strategy, likely due to distortions and artifacts introduced by the model during the first iteration.

\subsection{Team \emph{ZFTurbo} \anonym{(Roman Solovyev, Alexander Stempkovskiy, Tatiana Habruseva)}}
\label{sec:team_zfturbo}

\alter{We describe below our system for leaderboard \emph{Standard}, which ranked \nth{2} place, with 9.26dB SDR (mean over all instruments).}

\paragraph{\alter{Ensemble Models for Better Separation}}
\change{
Nowadays the open-source community has already collected several models for music source separation and made them available for usage~\citep{stoter2019open,hdemucs,rouard2023hybrid,spleeter2020}.
These models perform relatively well without the need for training them further.
We argue that, in order to create a model for source separation able to score well in a competition, we do not need to train new models and can instead reuse existing ones and combine them to increase their performance.
}

\change{
The practice of creating ensembles of models is not new~\citep{uhlich2017improving}, but no clear guidelines on how to create an effective ensemble exist.
Typically, the simplest solution is to find an ensemble of models that performs well on a small selection of songs and assume that the same performance will be achieved on all other songs.
Our approach aims to make the process of selecting and combining models easier, faster and more effective.
}

\change{
For this purpose, we created an open benchmark~\citep{solovyev2023benchmarks} for sound demixing that we use to run objective evaluations using a standardized interface.
This system allows us to quickly add more models and more evaluation data, with the final objective of finding the best ensemble of models available.
This evaluation system is available online\endnote{\url{https://mvsep.com/quality\_checker/}}.
At the time of writing, our evaluation system features two datasets: a \emph{synthetic} one, where small segments of vocals and instrumental songs are mixed together, even if they are not musically coherent, and a realistic one, called \emph{MultiSong dataset}, which comprises 100 one-minute tracks of varying genre.
Although the data at our disposal for the evaluation still represent a relatively small subset of the possible scenarios a separation model will encounter during inference, our approach is already a major improvement with respect to the practice of manually selecting ensemble weights based on subjective listening of a small selection of songs.
}

\paragraph{Details of the Proposed \alter{Ensemble}}

\change{
In the following, we describe the ensemble model that we submitted to the challenge.
We found the coefficients to ensemble the individual models using our evaluation system.
It is important to note how we use different combinations of models for extracting different sources.
The best ensemble for one instrument likely is not the best ensemble for another instrument.
This requires us to run independent evaluations on every single source to be extracted.
}

To \change{extract} the vocals, we \change{combine} three pretrained models: UVR-MDX1\endnote{Checkpoint ``Kim\_Vocal\_1.onnx'' available at \url{https://github.com/TRvlvr/model_repo/releases/download/all_public_uvr_models/Kim_Vocal_1.onnx}}, UVR-MDX2\endnote{Checkpoint ``UVR--MDX--NET--Inst\_HQ\_2.onnx'' available at \url{https://github.com/TRvlvr/model_repo/releases/download/all_public_uvr_models/UVR-MDX-NET-Inst_HQ_2.onnx}} \change{(these models are part of the} Ultimate Vocal Remover project\endnote{\url{https://github.com/Anjok07/ultimatevocalremovergui}}) and HTDemucs~\citep{rouard2023hybrid} \change{(we use the fine-tuned version available on GitHub}\endnote{Checkpoint \emph{htdemucs\_ft} available at \url{https://github.com/facebookresearch/demucs}}).
\change{Please note that} UVR-MDX2 is a \change{model trained to estimate the instrumental part of a song (i.e., to remove the vocals), so we subtract its output from the original mixture to obtain an estimate for the vocals.
HTDemucs allows \alter{one} to estimate the separated signals several times by shifting the input mixture over time by a small random amount.
The estimations are then aggregated after compensating for the shift.
We found that increasing the number of estimations (each with a different shift) increases the performance, with the improvement becoming negligible for very high values.
Moreover, HTDemucs splits the input signal in overlapping windows, runs independent predictions on all of them and combines the estimations at the end.
We can choose how much overlap to have between consecutive windows.
We found that maximizing the overlap increases the separation quality, at the cost of slower inference.
Also, for high overlap values the gain in performance can become negligible.
We report the improvements we observed using these two features in Table~\ref{tab:zfturbo_1_shift_overlap}.
Additionally, we found beneficial to separate the song twice: once using the original signal, once inverting its phase (this is simply achieved by multiplying the waveform by $-1$).
The two estimations are then averaged together after compensating for the change in phase.
The outputs of UVR-MDX1, UVR-MDX2 and HTDemucs are combined with a weighted sum of the time-domain signals.
}

\change{
\begin{table}[ht]
\centering
\scriptsize
\renewcommand{\arraystretch}{1.2}
\begin{tabular}{ccccccc}
\toprule
& & \multicolumn{4}{c}{\textbf{Global SDR} (dB)} \\
\cline{3-7}
Shifts & Overlap Ratio & Mean & Bass & Drums & Other & Vocals \\
\midrule
2  & 0.5  & 9.43 & 12.15 & 11.35 & 5.81 & 8.40 \\
4  & 0.75 & 9.47 & 12.22 & 11.40 & 5.84 & 8.41 \\
1  & 0.95 & 9.48 & 12.24 & \textbf{11.41} & 5.84 & \textbf{8.43} \\
10 & 0.95 & \textbf{9.49} & \textbf{12.25} & \textbf{11.41} & \textbf{5.85} & \textbf{8.43} \\
\bottomrule
\end{tabular}
\vspace{0.1cm}
\caption{(Team \emph{ZFTurbo}) Separation performance varying the number of shifts and overlap during the inference of HTDemucs. Increasing both lead to higher performance, with marginal improvements for very high parameter values.}
\label{tab:zfturbo_1_shift_overlap}
\end{table}
}

\change{
In order to extract the remaining three instruments, we found beneficial to compute the instrumental track and use it as input of the separation models instead of the original mixture.
We compute the instrumental track by subtracting the estimation of the vocals from the original song.
Intuitively, this simplifies the task of the remaining models, as the vocals do not need be removed anymore.
We use four different variants of HTDemucs among those available on GitHub\endnote{\url{https://github.com/facebookresearch/demucs}} (\textit{demucs\_ft}, \textit{demucs}, \textit{demucs\_6s}, \textit{demucs\_mmi}) and produce one estimate of each instrument for each of them. Please note that we use the phase inversion method described above for each model.
We then combine the four outputs (one per model) using specific weights for each instrument.
To produce the final estimates, we further combine the estimation of models for different instruments, to further make sure that the estimations are as robust/stable as possible.
}

\change{
Table~\ref{tab:zfturbo_1} reports the SDR scores for our final model ensemble on MDXDB21 (the test set used in the challenge) and on our MultiSong dataset.
For all details about our approach}, the source code is publicly available on GitHub\endnote{\url{https://github.com/ZFTurbo/MVSEP-MDX23-music-separation-model}}.

\begin{table}[ht]
\centering
\scriptsize
\renewcommand{\arraystretch}{1.2}
\begin{tabular}{lccccc}
\toprule
& \multicolumn{5}{c}{\textbf{Global SDR} (dB)} \\
\cline{2-6}
Dataset & Mean & Bass & Drums & Other & Vocals \\
\midrule
MultiSong MVSep & \textbf{10.11} & \textbf{12.68} & \textbf{11.68} & 6.67 & 9.62 \\
MDXDB21 (18 songs) & 9.41 & 9.87 & 9.52 & \textbf{7.43} & \textbf{10.81} \\
MDXDB21 (27 songs) & 9.25 & 9.94 & 9.53 & 7.05 & 10.51 \\
\bottomrule
\end{tabular}
\vspace{0.1cm}
\caption{(Team \emph{ZFTurbo}) SDR scores for the final ensemble on our MultiSong dataset~\citep{solovyev2023benchmarks} and on MDXDB21. We report separately the scores visible during the competition (only on 18 songs) and at the end (on 27 songs).}
\label{tab:zfturbo_1}
\end{table}

\subsection{Team \emph{subatomicseer} \anonym{(Nabarun Goswami, Tatsuya Harada)}}
\label{sec:team_subatomicseer}

\alter{We focus on our approaches for robust MSS, submitted to leaderboards \emph{LabelNoise} and \emph{Bleeding}, which ranked \nth{2} and \nth{3} place respectively.
Please note that we use different models for the two leaderboards.
Then, we briefly present the system we submitted to the leaderboard \emph{Standard}, which ranked \nth{7} place.}

\paragraph{\alter{Discrete Wavelet Transform in MSS}}

\change{
Signals of different musical instruments have distinct features at several resolutions.
We argue that exploiting the resolutions of such features can be helpful in carrying out tasks such as source separation.
For this reason, we propose to take advantage of the} multi-resolution analysis capabilities of \change{the} discrete wavelet transform (DWT) and \change{introduce} two new models for source separation \change{that make use of it.
Moreover, we} propose a noise-robust training scheme to \change{train models for robust MSS.
}

\change{
The existing HTDemucs~\citep{rouard2023hybrid} processes the signals in two separate branches, one for time-domain waveforms and one for spectrograms.
The two branches communicate using a cross-transformer.
We propose to add a third branch, which operates on the DWT of the signals.
This branch is based on a stack of} encoders, decoders and cross-transformers.
\change{The branch that processes spectrograms acts as a} residual bridge between the \change{branch that processes the DWT and the one that processes time-domain waveforms.
We call this architecture \emph{Wavelet HTDemucs} (WHTDemucs).
}

\change{
We propose also a second model, based on a simpler architecture, which we name \emph{DWT Transformer UNet} (DTUNet).
It is based on two separate branches: one for the time-domain waveform, one for the DTW.
A single cross-transformer allows communication between the two branches.
The outputs of the two branches are element-wise summed.
We also apply to the final output} a source-independent post-filter\change{, composed of a stack of three convolutional layers with GELU activations.
}

\paragraph{Robust Training}

\change{
To train models for robust MSS, we use the following strategy.}
\alter{From the start of the training, in order to fix the source permutation at the output of the model, we employ a reconstruction objective that is not permutation invariant.
We employ the L1 loss, which has been shown to work well for supervised audio source separation training~\citep{Braun2020ACV,rouard2023hybrid,luo2023music}.}
\change{Throughout the training, we rely on the} Mixture Consistency loss~\citep{8682783} \change{to prevent the model from collapsing to outputting silence.
}

\change{
Additionally, we employ the} unsupervised MixIT loss~\citep{wisdom2020unsupervised}\change{.
Having four sources, we group them in two pairs (randomly). The signals in each pair are mixed together. These two mixes represent the separation targets, given the original mix as input.
Moreover, we incorporate} the Mean Teacher loss~\citep{tarvainen2017mean}\change{, where a student is trained to estimate the output of a teacher model (in our case, the} exponential moving average (EMA) \change{of the model parameters).
We experiment with two variants of the Mean Teacher loss.
In version 1 (V1) we compute the mean teacher target by mixing the outputs of the EMA model for each noisy input stem.
In version 2 (V2) we only use the output of the EMA model on the input mixture.
When using V1, we apply it after 30k iterations; when using V2, we apply it after 75k iterations, using a smaller batch size.
}

\change{
In order to address the task of leaderboard \emph{LabelNoise}, we also design a method to remove the noisy stems from the training data.
We train a DTUNet model using V2 of the Mean Teacher loss.
We then separate all the individual stems in the training data and compute the SDR between them and the model output corresponding to the stem label.
If the SDR value is higher than 9dB, we consider the stem potentially clean.
We then} manually check a subset of the stems \change{whose separation exceeds 9dB and only keep those effectively clean.
Finally, we train a BSRNN model~\citep{luo2023music} (one per instrument) on this clean data.
The final submission is an ensemble of a WHTDemucs (trained on the full \emph{SDXDB23\_LabelNoise} with V1), a DTUNet (trained on the full \emph{SDXDB23\_LabelNoise} with V2) and a BSRNN (trained on the clean data).
We select the ensemble weights based on the performance of the individual models on the leaderboard and by manually inspecting the separations.
Table~\ref{tab:subatomicseer_LB-A} reports the performance of the two proposed models and of the final ensemble for the Label Noise task.
}

\change{
\begin{table}[ht]
\centering
\scriptsize
\renewcommand{\arraystretch}{1.2}
\begin{tabular}{lccccc}
\toprule
& \multicolumn{5}{c}{\textbf{Global SDR} (dB)} \\
\cline{2-6}
Model (Mean Teacher loss) & Mean & Bass & Drums & Other & Vocals \\
\midrule
WHTDemucs (V1) & 5.93 & 6.41 & 5.73 & 4.42 & 7.17 \\
DTUNet (V2) & 5.93 & 5.84 & 6.71 & 4.10 & 7.08 \\ 
Blend & \textbf{6.60} & \textbf{6.70} & \textbf{7.03} & \textbf{4.61} & \textbf{8.07} \\
\bottomrule
\end{tabular}
\vspace{0.1cm}
\caption{(Team \emph{subatomicseer}) Our scores on the \emph{LabelNoise} leaderboard.}
\label{tab:subatomicseer_LB-A}  
\end{table}
}

\change{
\alter{Manually selecting a clean subset of the dataset as described above could only be done} for the training set with label noise, since all the stems in \emph{SDXDB23\_Bleeding} were corrupted.
Therefore, our submission for leaderboard \emph{Bleeding} comprises an ensemble of a WHTDemucs (trained on \emph{SDXDB23\_Bleeding} with V1) and a DTUNet (trained on \emph{SDXDB23\_Bleeding} with V2).
We select the ensemble weights based on the performance of the individual models in the leaderboard.
Table~\ref{tab:subatomicseer_LB-B} reports the performance of the two proposed models and of the final ensemble for the Bleeding task.
}

\change{
\begin{table}[ht]
\centering
\scriptsize
\renewcommand{\arraystretch}{1.2}
\begin{tabular}{lccccc}
\toprule
& \multicolumn{5}{c}{\textbf{Global SDR} (dB)} \\
\cline{2-6}
Model (Mean Teacher) & Mean & Bass & Drums & Other & Vocals \\
\midrule
WHTDemucs (V1) & 5.86 & 5.90 & 5.61 & \textbf{4.68} & 7.25 \\ 
DTUNet (V2) & 5.62 & 5.37 & 6.18 & 3.92 & 7.00 \\ 
Blend & \textbf{6.31} & \textbf{6.33} & \textbf{6.86} & 4.59 & \textbf{7.47} \\
\bottomrule
\end{tabular}
\vspace{0.1cm}
\caption{(Team \emph{subatomicseer}) Our scores on the \emph{Bleeding} leaderboard. }
\label{tab:subatomicseer_LB-B}  
\end{table}
}

\paragraph{\emph{Standard} Leaderboard}

\change{
For the leaderboard \emph{Standard}, we train a DTUNet and a BSRNN on all the data at our disposal: MUSDB18~\citep{rafii2017musdb18}, MedleyDB-V1~\citep{bittner2014medleydb}, and stems from our private collection, for a total of 347 songs.
Our final submission is an ensemble of those two models and HTDemucs (the default model available on GitHub\endnote{\url{https://github.com/facebookresearch/demucs}}).
We select the ensemble weights that maximize the score on our validation set.
Table~\ref{tab:subatomicseer_LB-C-local} reports the results of the three individual models on our validation set.
}

\begin{table}[ht]
\centering
\scriptsize
\renewcommand{\arraystretch}{1.2}
\begin{tabular}{lccccc}
\toprule
& \multicolumn{5}{c}{\textbf{Global SDR} (dB)} \\
\cline{2-6}
Model (Training Songs) & Mean & Bass & Drums & Other & Vocals \\
\midrule
DTUNet (347) & 8.79 & 8.75 & 10.65 & 6.76 & 8.99 \\ 
BSRNN (347) & 8.65 & 8.06 & \textbf{10.80} & 6.38 & \textbf{9.37} \\ 
HTDemucs (800) & \textbf{9.19} & \textbf{9.68} & 10.76 & \textbf{7.17} & 9.15 \\
\bottomrule
\end{tabular}
\vspace{0.1cm}
\caption{(Team \emph{subatomicseer}) Performance of the individual models of our ensemble on our validation set. Please note that HTDemucs is trained with more data than our internal models.}
\label{tab:subatomicseer_LB-C-local}  
\end{table}

\subsection{Team \emph{CCOM} \anonym{(Yuanliang Dong, Xinran Zhang, Jiafeng Liu)}}
\label{sec:team_CCOM}

\alter{We focus on our approach for leaderboard \emph{LabelNoise}, which ranked \nth{1} place, with 7.46dB SDR (mean over all instruments).}

\paragraph{Robust Training}
\change{
Our method to train a model for robust MSS is composed of two steps: first, we train the model by making the loss robust to noisy ground truth stems, then we use the trained model to detect and exclude noisy stems in the training data and use the new dataset to train a better model.
}

\change{
First, we make the training loss robust to noisy stems using the \emph{loss truncation} technique.}
The idea of \emph{loss truncation} was introduced by~\citet{kang_improved_natural_language}.
Suppose an oracle model exists that \change{can perfectly separate each stem (i.e., with infinite SDR).
With this model,} correct samples \change{in the training data would have zero loss, while noisy samples would have a loss greater than zero.
We can consider} this oracle model a perfect classifier \change{for the classes \emph{clean labels} and \emph{noisy labels}, using the (quantile of the) loss as classification criterion.
In practice, in a batch of samples, we can sort the samples by their loss value in descending order, calculate some quantile of the losses as a threshold and discard the samples above such threshold.
Such \alter{an} oracle model does not exist in practice, and the rules of the challenge prevent us from using an existing source separation model trained on clean data as an approximation.
Therefore, we assume that a source separation model trained on noisy data is good enough at approximating the oracle method and train HTDemucs on \emph{SDXDB23\_LabelNoise} using \emph{loss truncation} from the beginning (i.e., the model being trained and the one approximating the oracle are the same).
}

\change{
In order to further improve the separation quality, we design a process to remove the stems with corrupted labels from the dataset.
We start from the assumption that the HTDemucs model trained with \emph{loss truncation} already has the ability to discriminate between a stem with a noisy label and one with a correct label.
We then use this model to separate every individual stem in the dataset into the four instruments and measure the energy of each estimate.
For example, by processing a stem labeled \emph{vocals}, we obtain four estimates (one per instrument).
If the stem is clean (i.e., if the label is correct), only the estimate of the vocals should have positive energy.
On the other hand, if the stem is not clean, some other estimate will have higher energy.
In practice, we compare the energy of the estimate corresponding to the stem label and, if it is at least 20dB higher than the energy of the other estimates, we consider the stem clean.
After this procedure, we create a cleaned dataset of 517 stems (out of 812 stems).
We repeat this process another time, by retraining HTDemucs on the cleaned dataset (without \emph{loss truncation}), and create a new set of 519 clean stems.
This amounts to excluding approximately 36\% of all the stems.
Finally, we use these 519 stems to fine-tune the previous HTDemucs and further improve the results.
Due to limitations in time and computational resources, we only conduct two iterations of this process.
In principle, with more iterations we could further clean the dataset and yield models that perform better.
}

\change{
In general, since we are using a model on the same data we trained it upon, the model could estimate the wrong separation as it might have \emph{memorized} the wrong stem during the training procedure.
We implicitly address this since the model processes individual stems, instead of mixtures (during training, the model always experienced mixtures at the input).
In addition to that, we choose to explicitly address this by changing the parameters for the augmentations between training and inference.}
During training, \change{we apply random pitch-shifting on the audio by} $\pm$ 2 semitones \change{and randomly change the tempo by up to} 12\%.
\change{Instead, during inference, we transpose the input stems} up by 6 semitones \change{and double the speed of the original signal.
}

\change{
We report in Table~\ref{tab:ccom} the results of our approach, including all intermediate steps of our procedure.
}

\begin{table}[ht]
\centering
\scriptsize
\renewcommand{\arraystretch}{1.2}
{
\changetable
\begin{tabular}{lccccc}
\toprule
& \multicolumn{4}{c}{\textbf{Global SDR} (dB)} \\
\cline{2-6}
Training Setup & Mean & Bass & Drums & Other & Vocals \\
\midrule
Baseline & 4.96 & 5.07 & 5.76 & 3.14 & 5.85 \\
With \emph{loss truncation} & 6.26 & 6.94 & 6.62 & 4.45 & 7.09 \\
With filtered data (\nth{1}) & 6.89 & 7.34 & 7.58 & 4.88 & 7.74 \\
With filtered data (\nth{2}) & \textbf{7.46} & \textbf{8.12} & \textbf{7.99} & \textbf{5.34} & \textbf{8.37} \\
\bottomrule
\end{tabular}
}
\vspace{0.1cm}
\caption{(Team \emph{CCOM}) Performance of HTDemucs using our approach. The baseline is trained on \emph{SDXDB23\_LabelNoise}, then we train a model using \emph{loss truncation} only. We use this model to filter the dataset (denoted with \nth{1} in the Table) and train a new model. Finally, we repeat the dataset filtering (denoted with \nth{2}) and fine-tune the model to obtain the best performance.}
\label{tab:ccom}  
\end{table}

\subsection{Team \emph{kuielab} \anonym{(Minseok Kim, Jun Hyung Lee)}}
\label{sec:team_kuielab}

\alter{We focus on our approaches for robust MSS, submitted to leaderboards \emph{LabelNoise} and \emph{Bleeding}, which ranked \nth{3} and \nth{1} place respectively.
Please note that we use different models for the two leaderboards.
Then, we briefly present the system we submitted to the leaderboard \emph{Standard}, which ranked \nth{4} place.}

\change{
Our participation in the previous edition of the challenge~\citep{mitsufuji2022music} resulted in the creation of an improved TFC-TDF-UNet model (named v2)~\citep{mdxnet}.
In the context of this edition of the challenge, we propose TFC-TDF-UNet v3, which achieves higher separation quality than its predecessor and reduces the inference time.
Additionally, we also propose a training strategy to handle errors in the training data, for models trained for robust MSS.
}

\paragraph{\alter{Improvements to the TFC-TDF-UNet Model}}
\change{
TFC-TDF-UNet is a music source separation model first introduced by~\citet{tdf}: it uses a UNet architecture that processes the complex spectrogram of the mixture and estimates the complex spectrogram of the source.
Unlike Open-Unmix~\citep{stoter2019open}, this model does not estimate a mask for the spectrogram, but directly estimates all components of the separated source.
The first version employed time-frequency convolutions (TFC) together with a time-distributed fully-connected (TDF) block as fundamental building blocks.
The TFC-TDF-UNet v2~\citep{mdxnet} was proposed in the context of the MDX'21 challenge, as a way to reduce the computational complexity of the first version, while achieving a higher separation quality.
}

\change{
The new version of TFC-TDF-UNet (v3), which we submitted to the SDX'23 challenge, features several architectural improvements.
First, the architecture of the model is now based on the ResUNet structure~\citep{Zhang2017RoadEB}, where a TDF block is added in every residual path.
We employ a channel-wise sub-band filter-bank~\citep{Liu2020ChannelwiseSI} while also increasing the size of the frequency dimension in our spectrograms, to improve the spectral resolution of the signals.
We train a single model that estimates all sources, instead of training one model for each source.
We found particularly effective to concatenate the input spectrogram to the model activation before the final convolutional layer.
We replace batch normalization and ReLU activations with instance normalization and GELU activations respectively.
We employ L2 on the time-domain waveform as training loss, instead of the L1 loss.
}

\paragraph{Robust Training}
\change{
For the models trained for robust MSS, we employ a simple strategy to make the training procedure less sensitive to the errors in the ground truth data, based on the idea of \emph{loss truncation}~\citep{kang_improved_natural_language}.
Please note how this approach is the same as that used by the CCOM team, described in Section~\ref{sec:team_CCOM}.
We assume that, in the presence of a relatively good model, an error in the ground truth data will generate a high loss value, even if the estimation made by the model is good. In this case, the gradients computed on this data will not necessarily improve the quality of the model, so we choose to exclude them.
In other words, when we compute the loss for a minibatch during training, we discard the samples with a higher loss value and only update the model parameters using the gradients for the samples with a low loss value.
This approach is sufficient when training on data corrupted with label noise, but not with bleeding, as bleeding occurs in all stems in the dataset.
In this case, we found it beneficial to discard not only samples along the batch dimension, but also along the time dimension. In other words, only the time instants of a sample with a low loss value contribute to the learning procedure.
}

\change{
For leaderboards \emph{LabelNoise} and \emph{Bleeding}, we trained TFC-TDF-UNet v3 only on \emph{SDXDB23\_LabelNoise} and \emph{SDXDB23\_Bleeding} respectively.
In both trainings, we use the strategy above to make the optimization robust to noisy stems.
Table~\ref{tab:kuielab_2} reports an ablation study on the strategy we propose to enable noise-robust training.
The results show that our strategy is an effective method to handle the errors in the training data, especially for label noise.
}

\begin{table}[ht]
\centering
\scriptsize
\renewcommand{\arraystretch}{1.2}
{
\changetable
\begin{tabular}{lcccccc}
\toprule
& & \multicolumn{5}{c}{\textbf{Global SDR} (dB)} \\
\cline{3-7}
Task & Loss Truncation & Mean & Bass & Drums & Other & Vocals \\
\midrule
\emph{LabelNoise} & No & 5.05 & 5.31 & 5.31 & 3.45 & 6.12 \\
\emph{LabelNoise} & Yes & \textbf{6.26} & \textbf{6.43} & \textbf{6.38} & \textbf{4.64} & \textbf{7.58} \\
\midrule
\emph{Bleeding} & No & 5.80 & 6.11 & 5.86 & 4.36 & 6.87 \\
\emph{Bleeding} & Yes & \textbf{6.22} & \textbf{6.58} & \textbf{6.20} & \textbf{4.69} & \textbf{7.41} \\
\bottomrule
\end{tabular}
}
\vspace{0.1cm}
\caption{(Team \emph{kuielab}) Ablation study on loss truncation. Please note that these are the scores of an individual TFC-TDF-UNet v3 model, not of the final ensemble.}
\label{tab:kuielab_2}  
\end{table}

\paragraph{\emph{Standard} Leaderboard}

\change{
For leaderboard \emph{Standard}, our final submission is an ensemble of TFC-TDF-UNet v3, HDemucs\endnote{\url{https://github.com/facebookresearch/demucs/tree/v3}} (commonly known as \emph{Demucs v3}) and HTDemucs (also known as \emph{Demucs v4}).
For all details on the models used in the submission, please refer to~\citet{sdx23kim}.
}

\change{
Table~\ref{tab:kuielab_1} shows a comparison of the new version of TFC-TDF-UNet (v3) with the previous model (v2). Not only do we achieve higher SDR score on all instruments, but we also increase the inference speed on GPU.
}

\begin{table}[ht]
\centering
\scriptsize
\renewcommand{\arraystretch}{1.2}
\begin{tabular}{ccccccc}
\toprule
& \multicolumn{5}{c}{\textbf{Global SDR} (dB)} & \\
\cline{2-6}
Model & Mean & Bass & Drums & Other & Vocals & Speed \\
\midrule
v2 & 7.03 & 6.85 & 6.87 & 5.44 & 8.96 & 12.8x \\
v3 & \textbf{7.90} & \textbf{7.36} & \textbf{8.81} & \textbf{6.19} & \textbf{9.22} & \textbf{15.0x} \\
\bottomrule
\end{tabular}
\vspace{0.1cm}
\caption{(Team \emph{kuielab}) Comparison of TFC-TDF-UNets v2 and v3 on the MUSDB18-HQ benchmark. \emph{Speed} denotes the relative GPU inference speed with respect to real-time on the challenge evaluation server.}
\label{tab:kuielab_1}  
\end{table}

\section{Listening Tests Results}
\label{sec:listening_results}
In this \change{s}ection, we report the results of the listening test.
\alter{Only the teams who achieved the three highest SDR scores in the leaderboard \emph{Standard} (\emph{SAMI-ByteDance}, \emph{ZFTurbo} and \emph{kimberley\_jensen}) participated in the listening test.
The best performing model of each team was used to produce the audio clips used in the test.
}

\change{The assessors were expected to complete at least 72 comparisons each. At the end of the test, they performed a total of 583 comparisons.}
Each comparison involved the separated outputs produced by two of the three models under evaluation and the original mixture to serve as reference.

First, we quantified the interactions of each assessor with the audio samples.
From the data gathered on the Moises.ai testing platform, we found that the assessors \change{spent on average} $27.95 \pm 17.88$ seconds \change{to perform one comparison}.
The assessors performed an average of $3.13 \pm 1.93$ switches between the segments of each comparison.

\begin{figure*}
    \centering
    \subfloat[Number of evaluations won by each model.]{\includegraphics[width=0.45\textwidth]{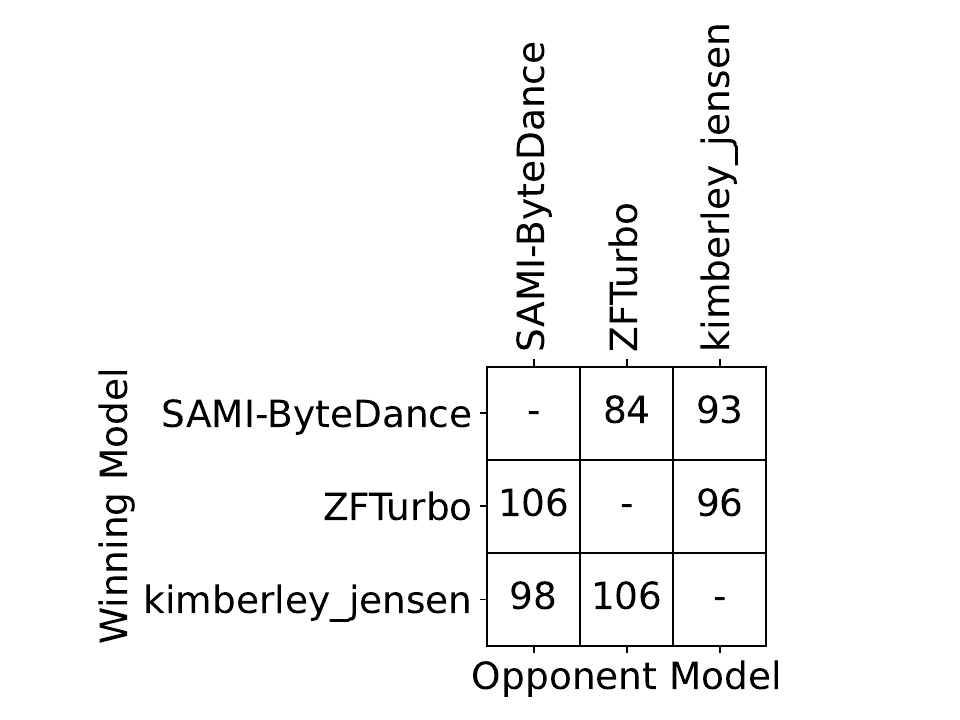}\label{fig:listening_test_results_global_absolute}}
    \hfill
    \subfloat[Normalized number of evaluations won by each model.]{\includegraphics[width=0.45\textwidth]{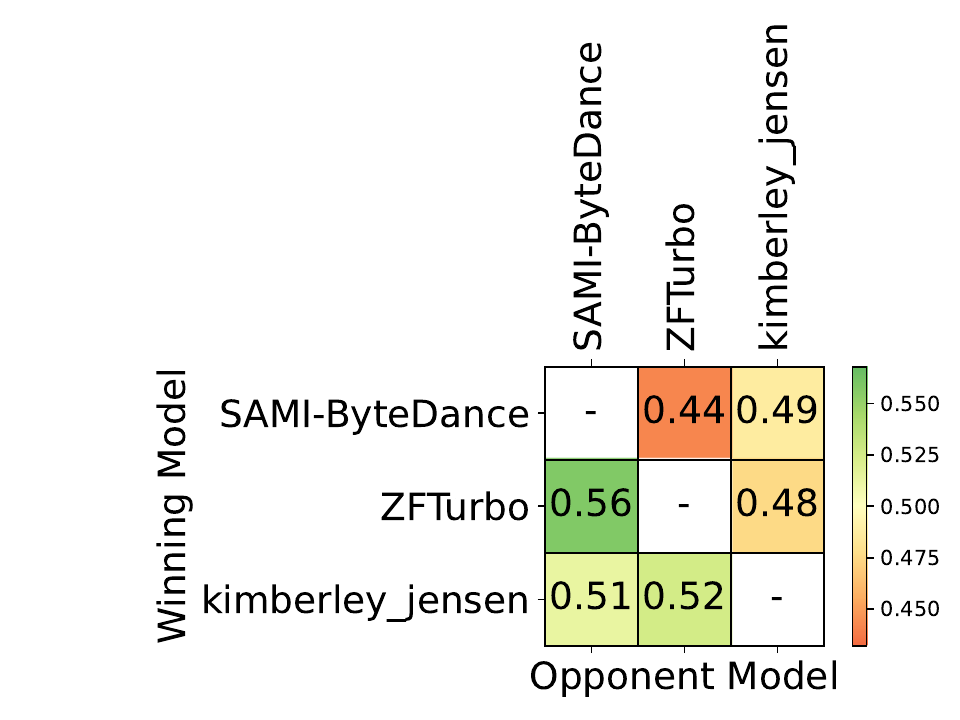}\label{fig:listening_test_results_global_relative}}
    \caption{Results of the listening test.}
    \label{fig:listening_test_results_global}
\end{figure*}

We report the global results of the listening test in Figure~\ref{fig:listening_test_results_global}.
The table in Figure~\ref{fig:listening_test_results_global_absolute} shows the number of times a model in each row won an evaluation against a model in each column.
We also report a second table (Figure~\ref{fig:listening_test_results_global_relative}) showing the same results normalized by the number of evaluations for each pair of models.

\begin{figure*}
    \centering
    \subfloat[Results for assessors in category \emph{Producer}.]{
        \includegraphics[width=0.45\textwidth]{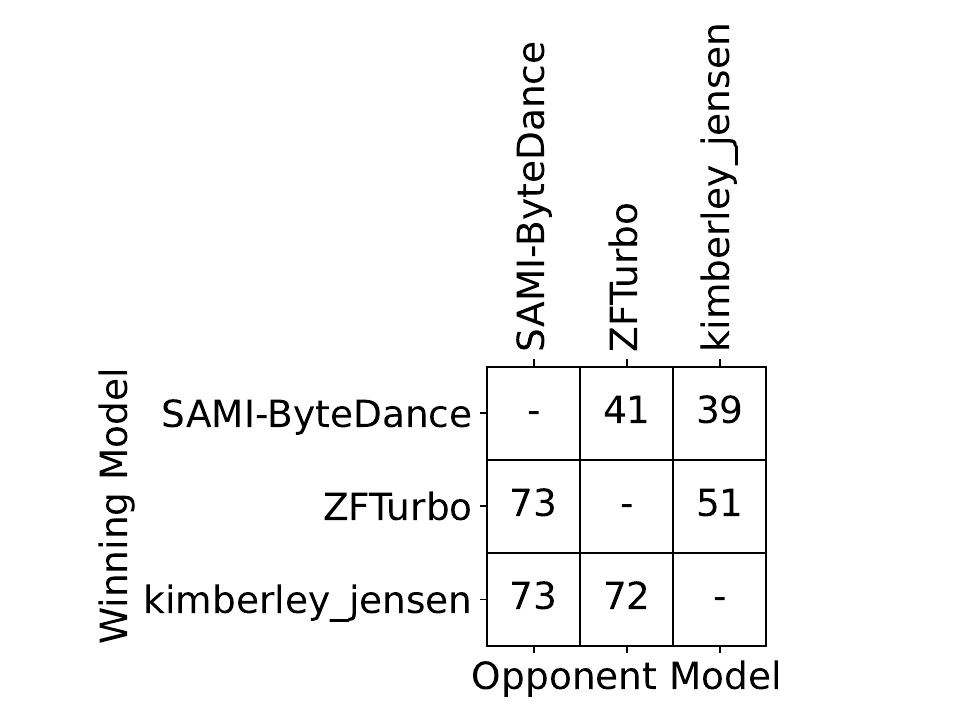}\label{fig:listening_test_results_categories_producer}
        \includegraphics[width=0.45\textwidth]{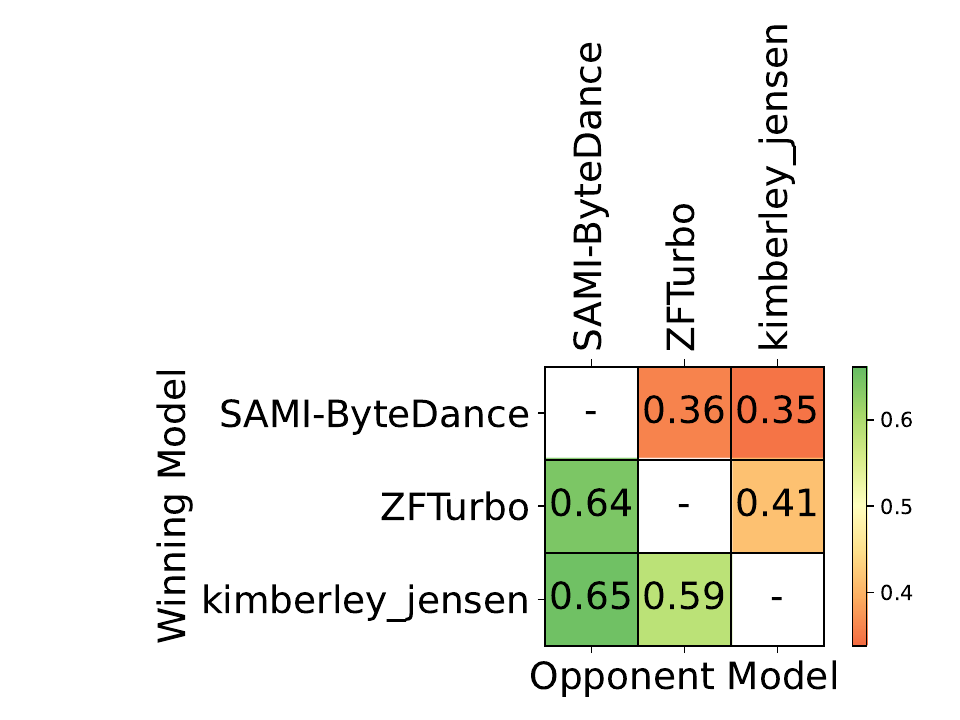}\label{fig:listening_test_results_categories_producer_relative}
    }
    \\
    \subfloat[Results for assessors in category \emph{Musician-Educator}.]{
        \includegraphics[width=0.45\textwidth]{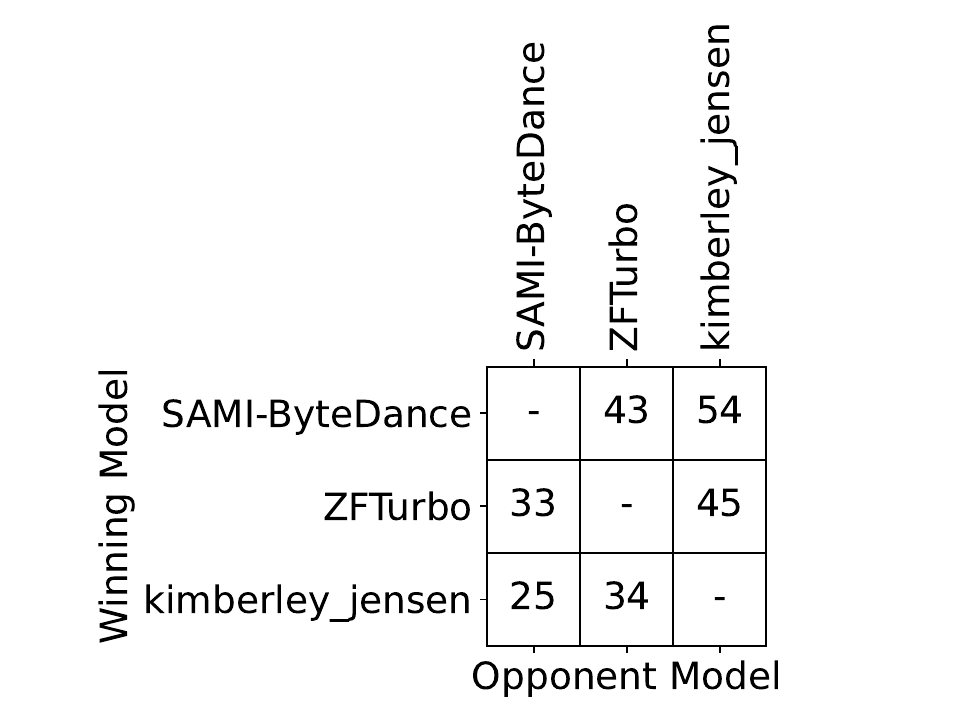}\label{fig:listening_test_results_categories_musician}
        \includegraphics[width=0.45\textwidth]{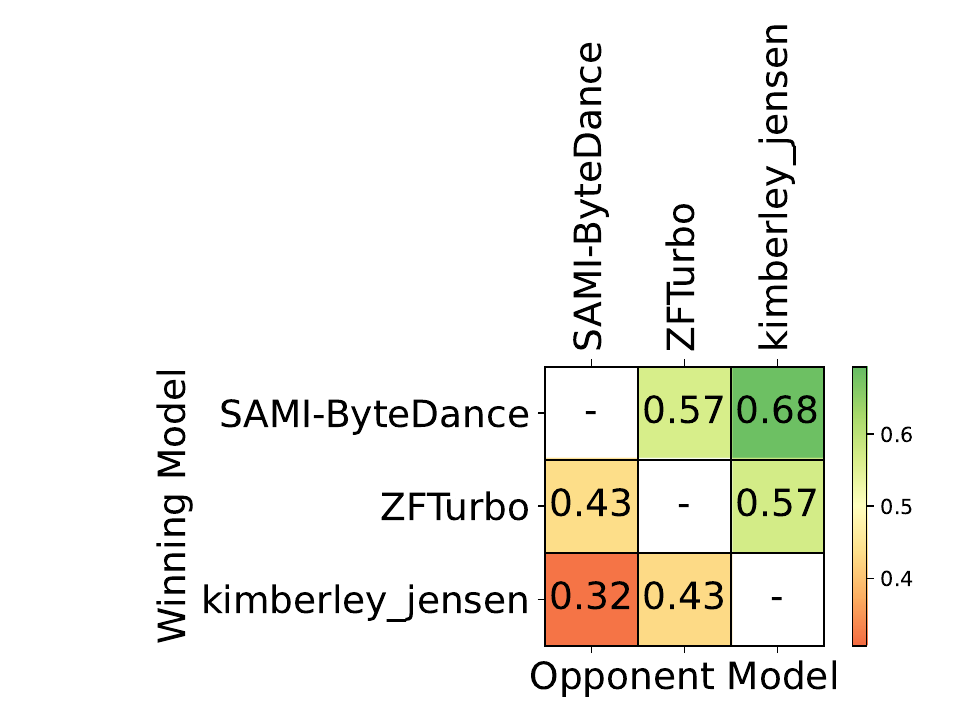}\label{fig:listening_test_results_categories_musician_relative}
    }
    \caption{Results of the listening test by assessor category.}
    \label{fig:listening_test_results_categories}
\end{figure*}

In order to detect potential biases as a result of our choice of assessors, we group them into two categories: \emph{Producer} and \emph{Musician-Educator} (we consider musicians and educators in the same category, since our assessor panel only includes one educator, who is also a performer).
We report the results of the listening test independently for the two categories in Figure~\ref{fig:listening_test_results_categories}.
Figure~\ref{fig:listening_test_results_categories_producer} shows the results for \emph{Producer}, while Figure~\ref{fig:listening_test_results_categories_musician} shows the results for \emph{Musician-Educator}.
We observe that assessors in the category \emph{Producer} showed a higher preference for \emph{kimberley\_jensen}, while those in \emph{Musician-Educator} preferred \emph{SAMI-ByteDance}.
\change{Given the limited number of assessors that took part in the listening test, it remains unclear whether this preference is the result of some statistical anomaly, or whether people with different backgrounds do really prefer one model over the other.}
\change{We leave this investigation to future work.}



\begin{figure*}
    \centering
    \subfloat[The evaluations performed on bass extraction.]{
        \includegraphics[width=0.25\textwidth, trim=10 10 10 10, clip]{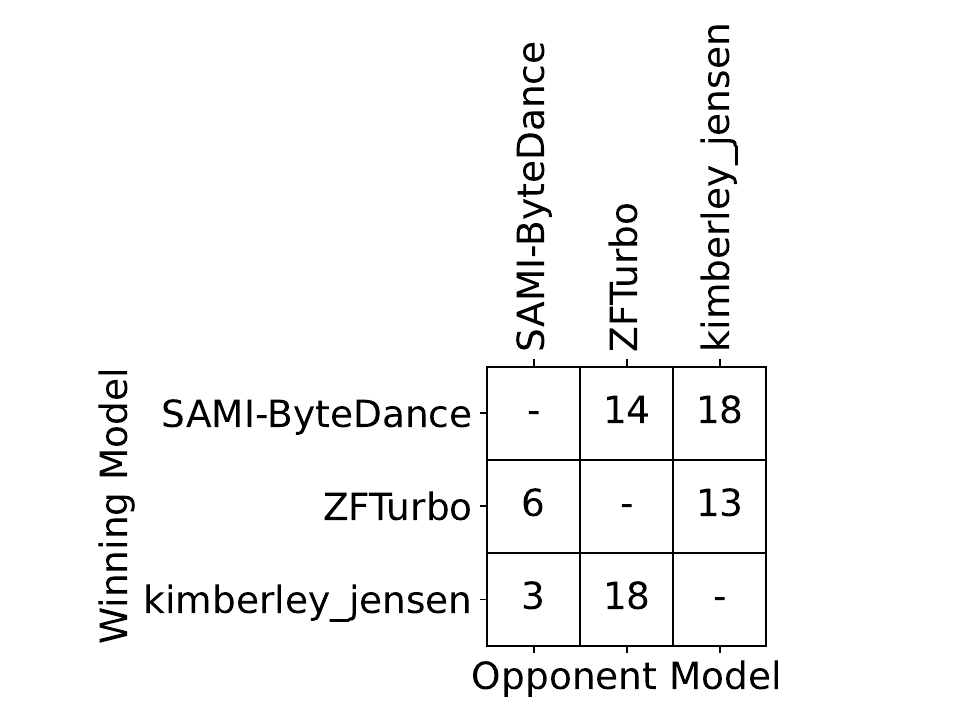}
        \includegraphics[width=0.25\textwidth, trim=10 10 10 10, clip]{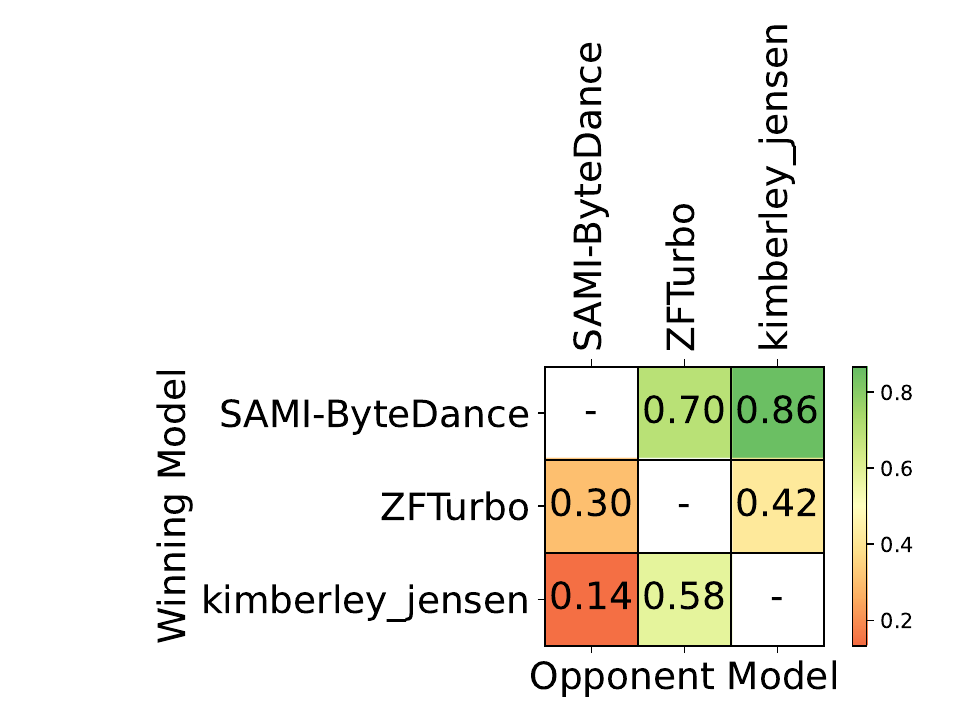}
    }
    \subfloat[The evaluations performed on bass removal.]{
        \includegraphics[width=0.25\textwidth, trim=10 10 10 10, clip]{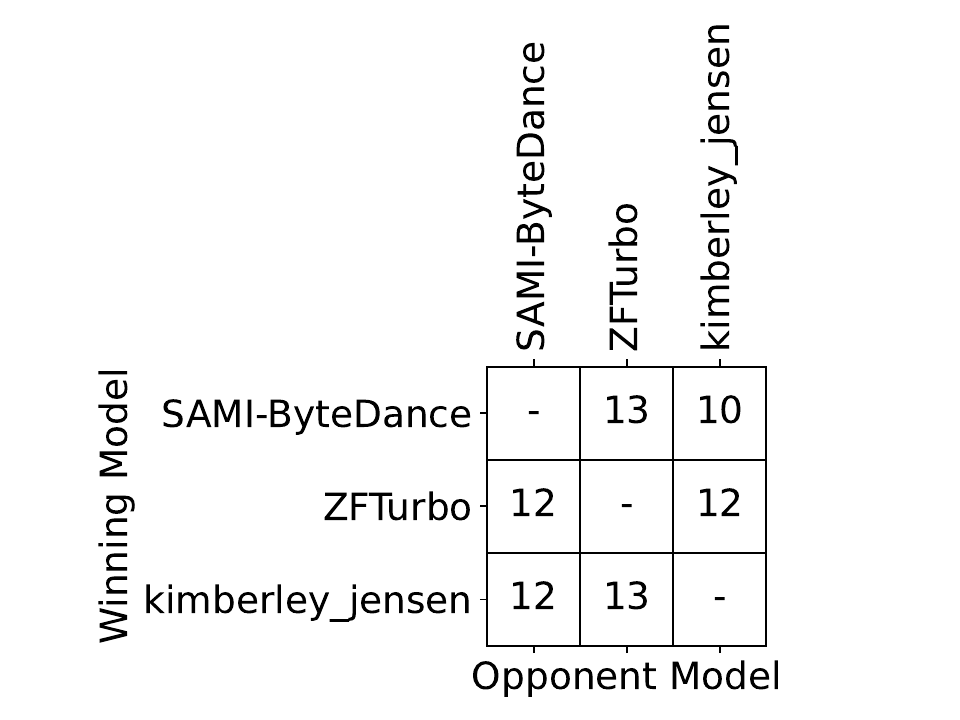}
        \includegraphics[width=0.25\textwidth, trim=10 10 10 10, clip]{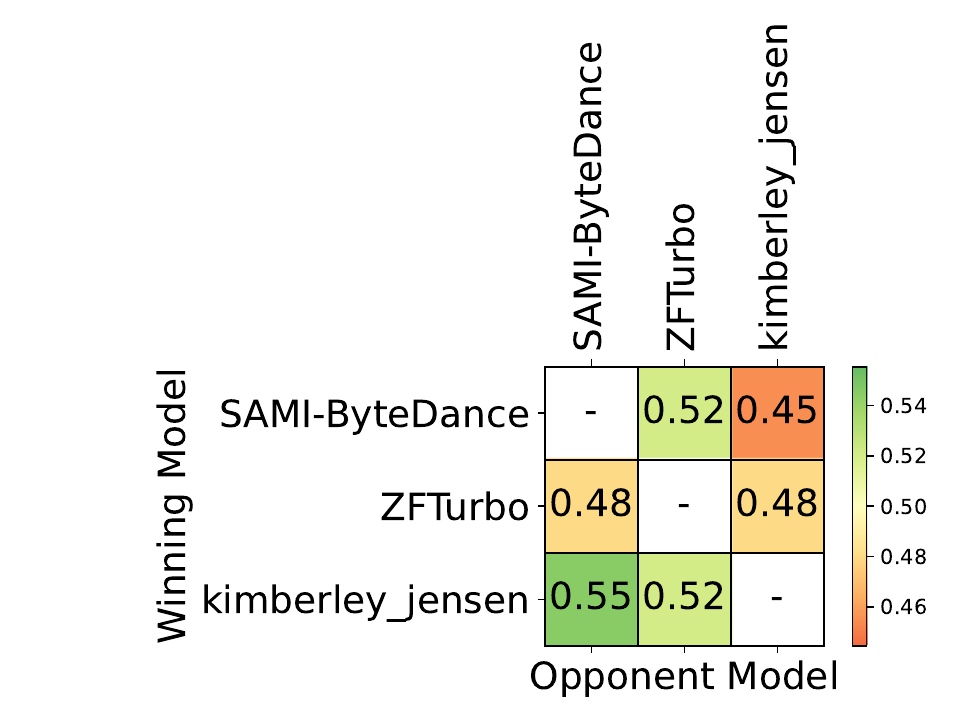}
    }
    \caption{Results of the listening test on bass removal and extraction.}
    \label{fig:listening_test_results_bass}
\end{figure*}

\begin{figure*}
    \subfloat[The evaluations performed on drum extraction.]{
        \includegraphics[width=0.25\textwidth, trim=10 10 10 10, clip]{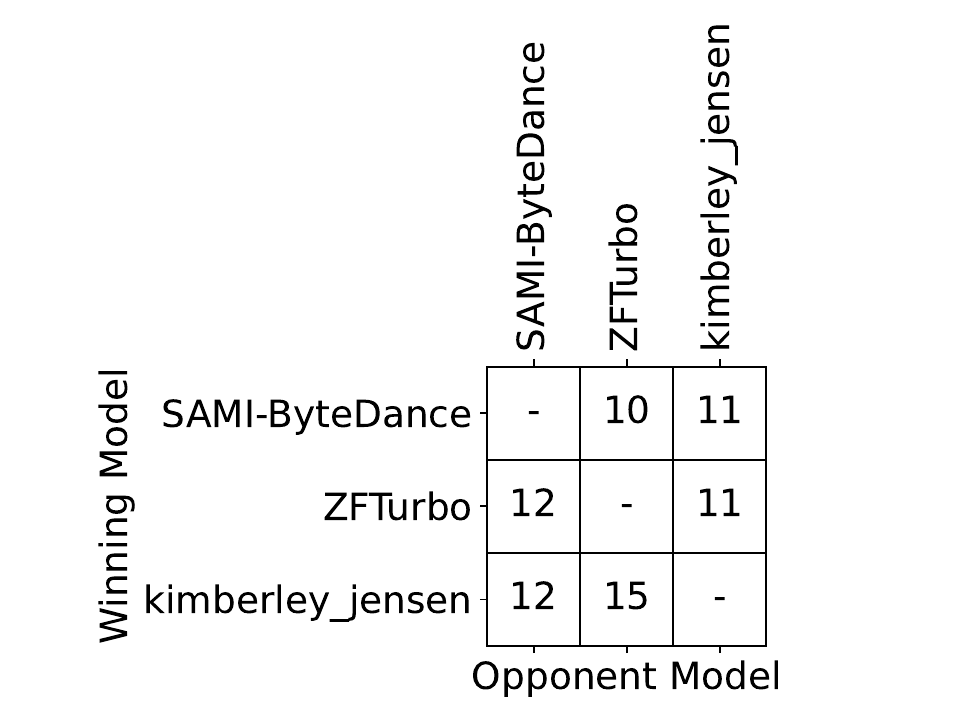}
        \includegraphics[width=0.25\textwidth, trim=10 10 10 10, clip]{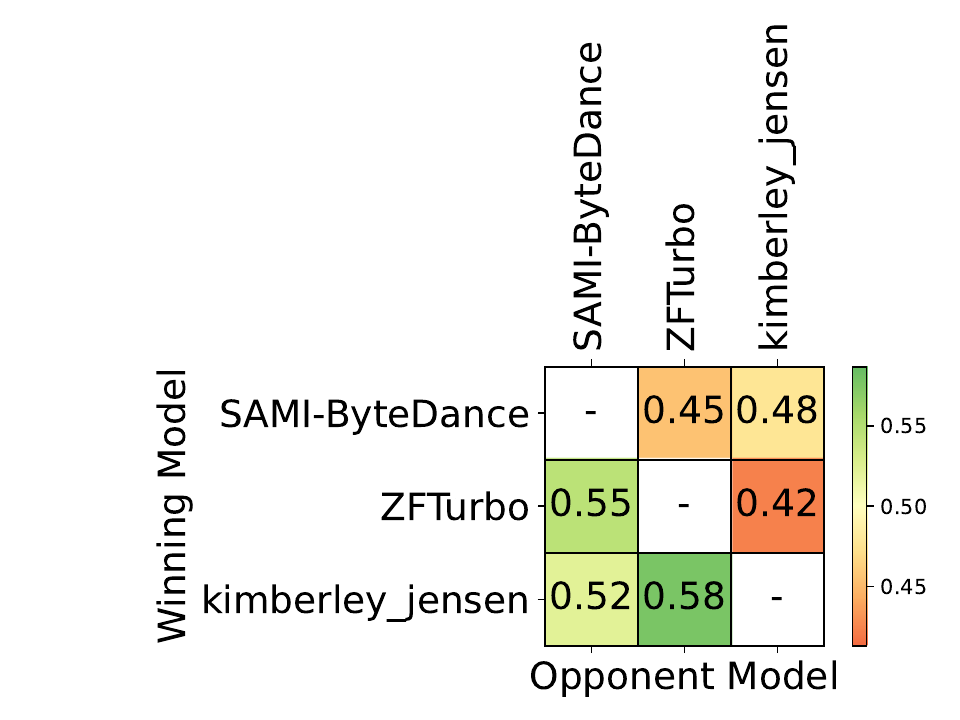}
    }
    \subfloat[The evaluations performed on drum removal.]{
        \includegraphics[width=0.25\textwidth, trim=10 10 10 10, clip]{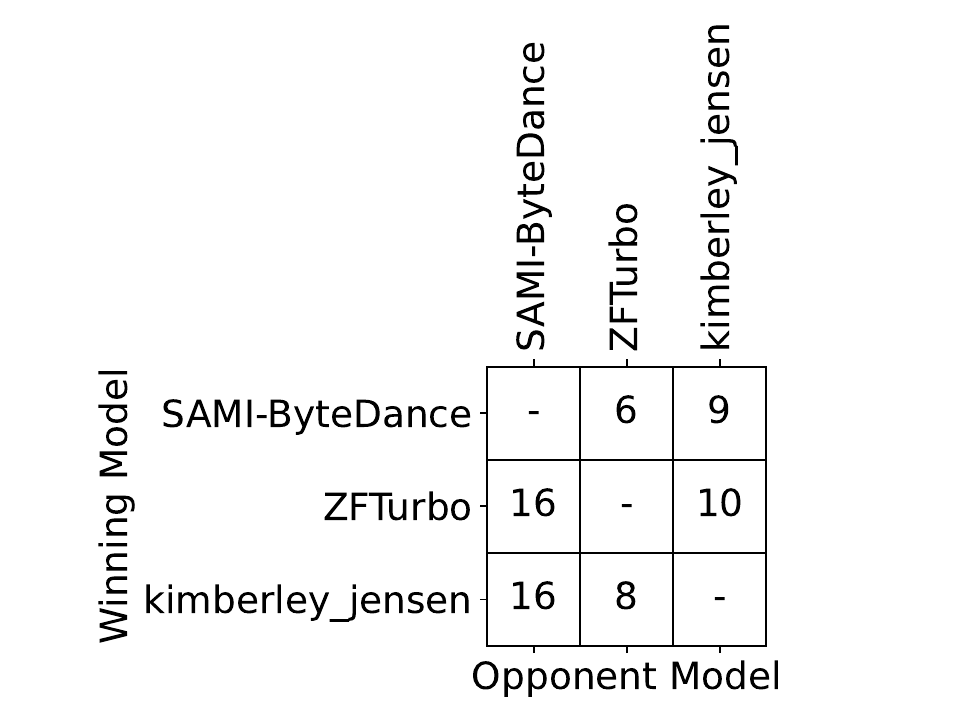}
        \includegraphics[width=0.25\textwidth, trim=10 10 10 10, clip]{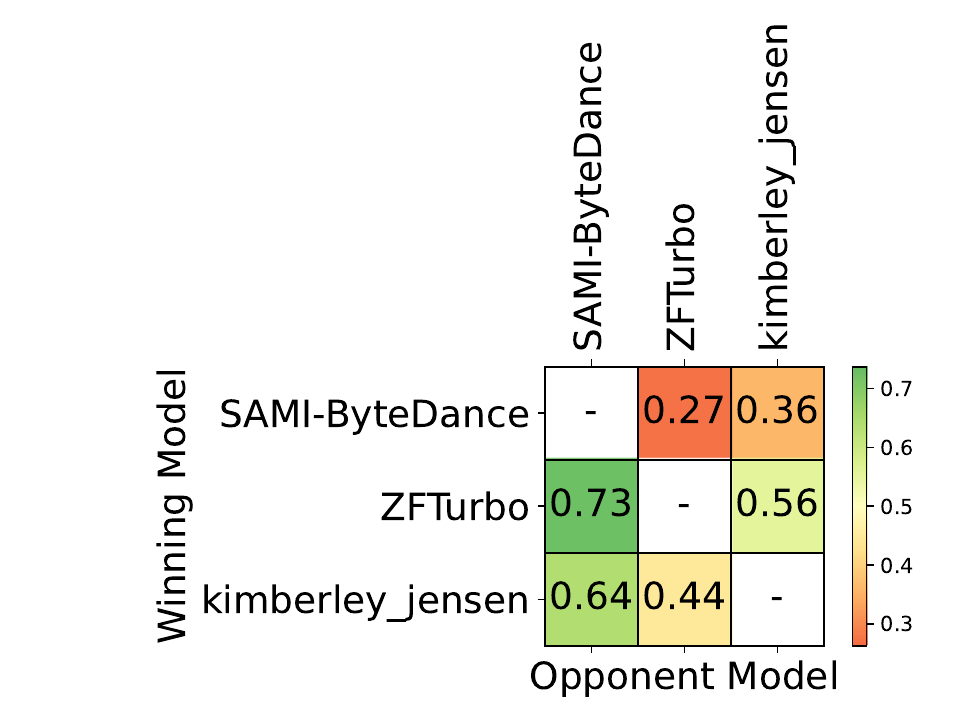}
    }
    \caption{Results of the listening test on drum removal and extraction.}
    \label{fig:listening_test_results_drums}
\end{figure*}

\begin{figure*}
    \subfloat[The evaluations performed on other extraction.]{
        \includegraphics[width=0.25\textwidth, trim=10 10 10 10, clip]{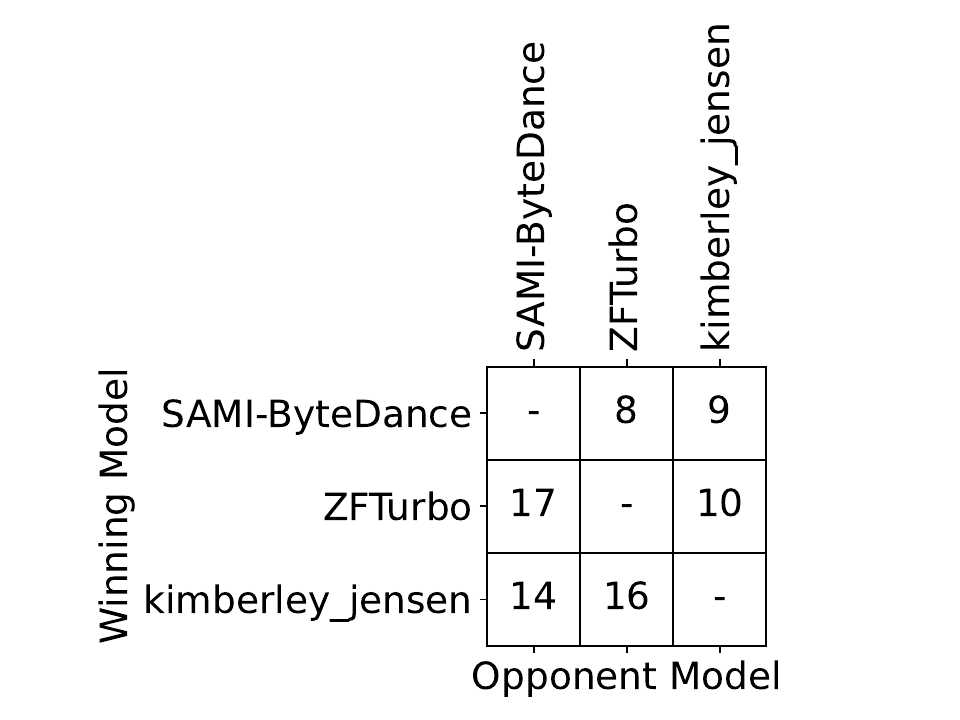}
        \includegraphics[width=0.25\textwidth, trim=10 10 10 10, clip]{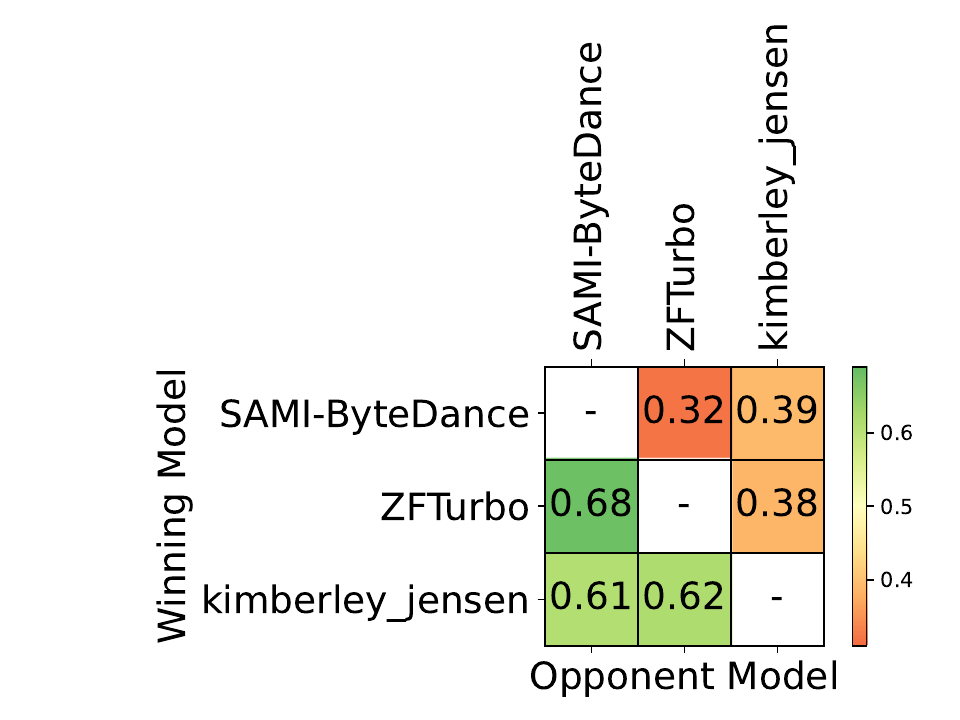}
    }
    \subfloat[The evaluations performed on other removal.]{
        \includegraphics[width=0.25\textwidth, trim=10 10 10 10, clip]{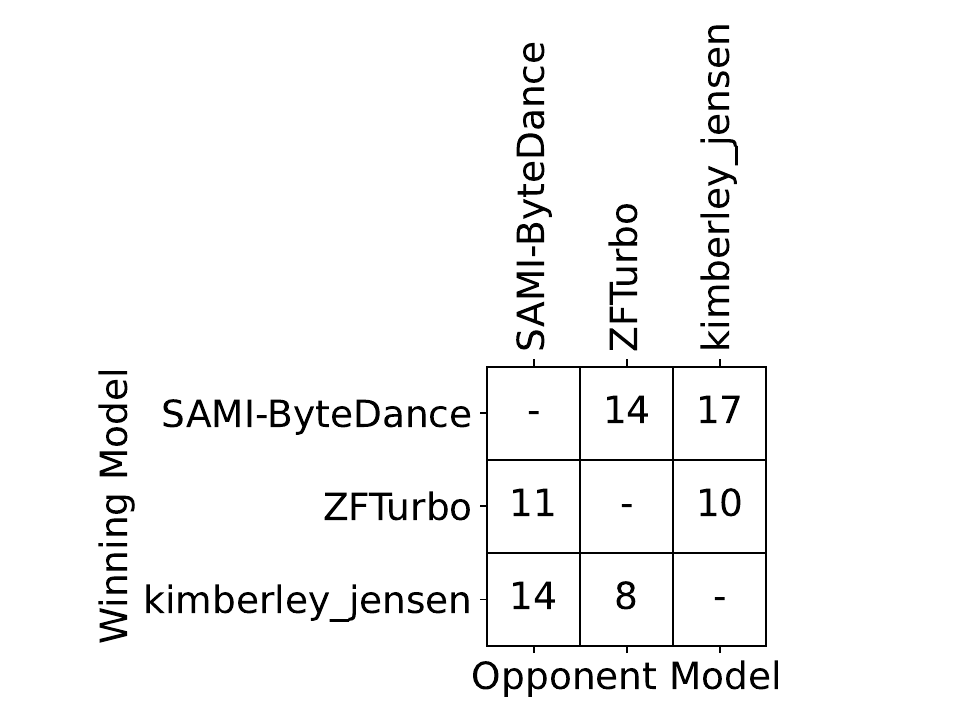}
        \includegraphics[width=0.25\textwidth, trim=10 10 10 10, clip]{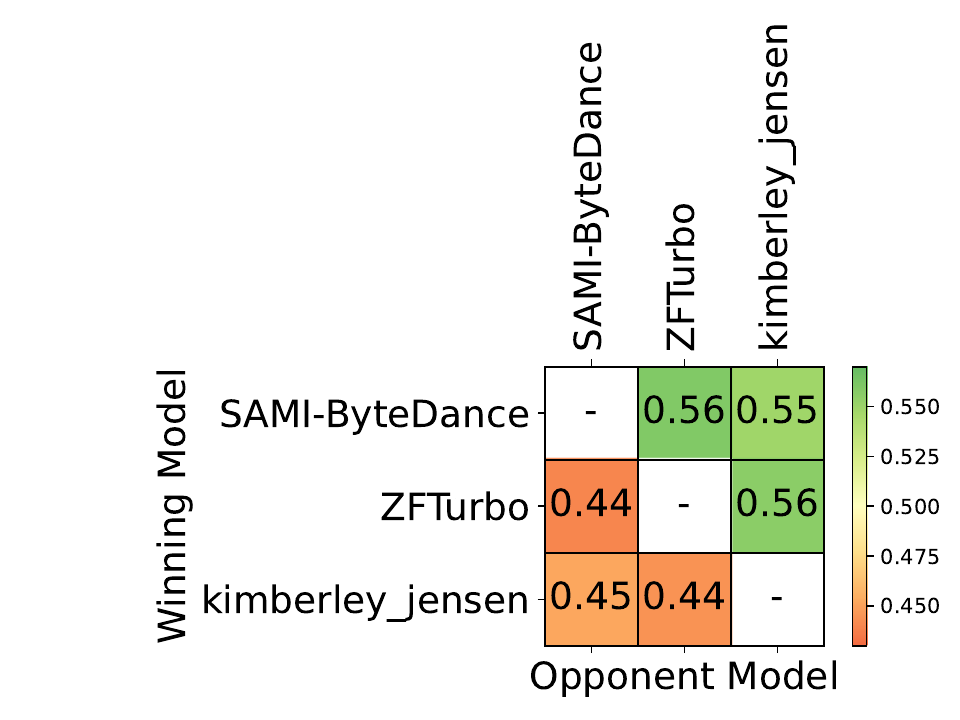}
    }
    \caption{Results of the listening test on other removal and extraction.}
    \label{fig:listening_test_results_other}
\end{figure*}

\begin{figure*}
    \subfloat[The evaluations performed on vocal extraction.]{
        \includegraphics[width=0.25\textwidth, trim=10 10 10 10, clip]{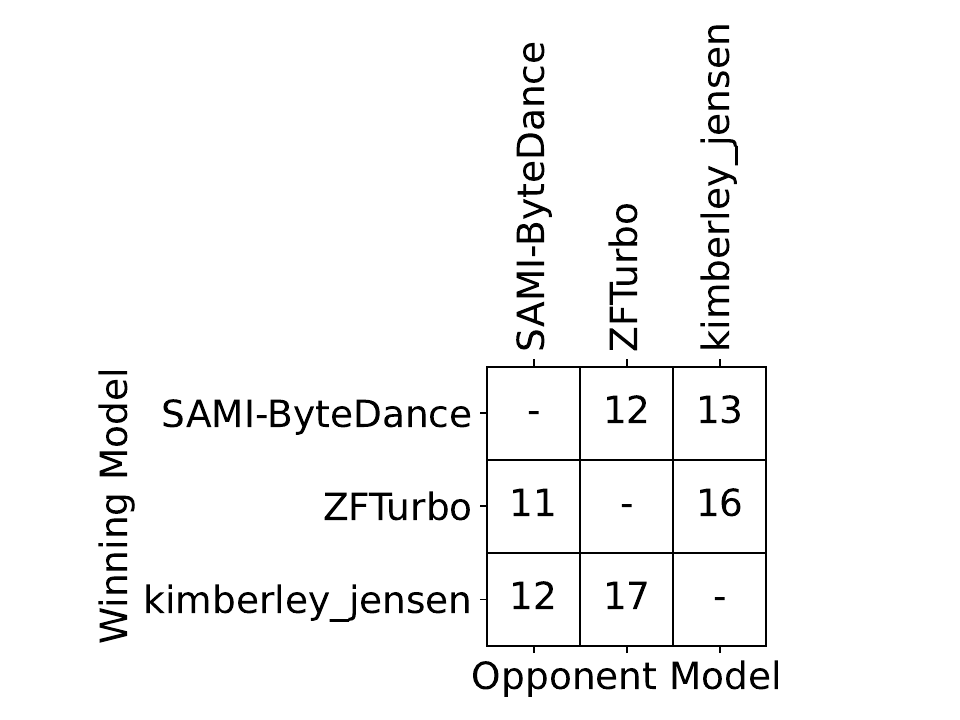}
        \includegraphics[width=0.25\textwidth, trim=10 10 10 10, clip]{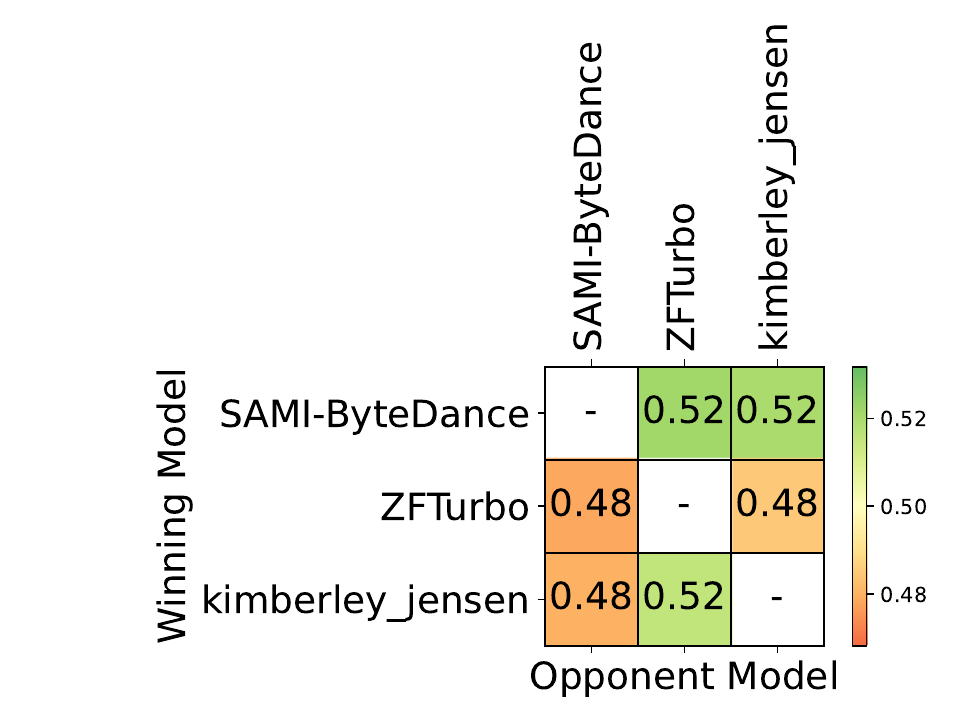}
    }
    \subfloat[The evaluations performed on vocal removal.]{
        \includegraphics[width=0.25\textwidth, trim=10 10 10 10, clip]{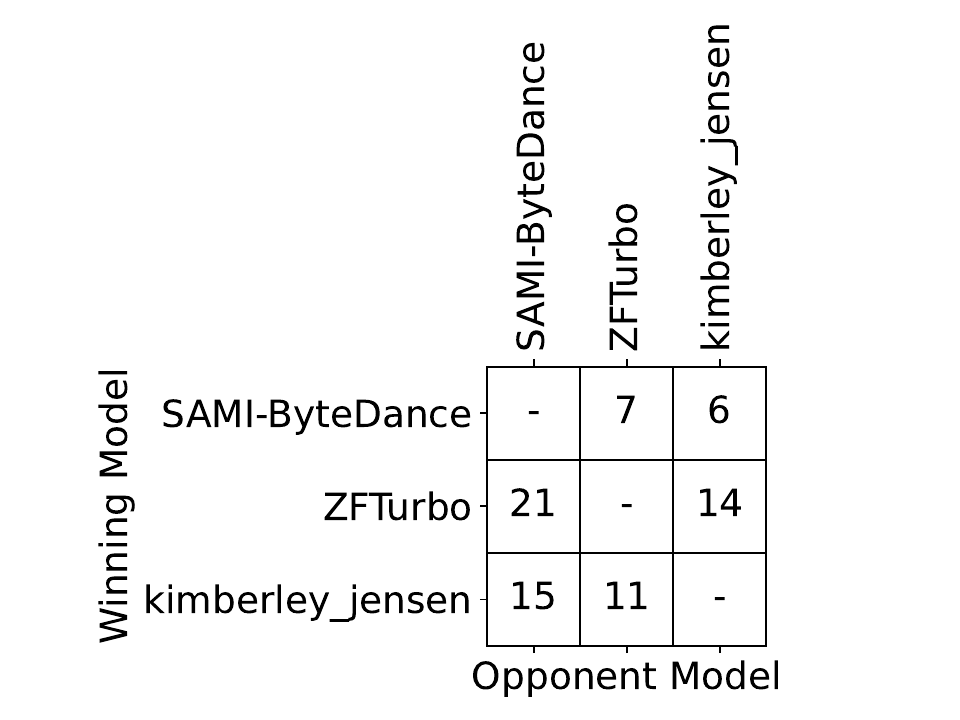}
        \includegraphics[width=0.25\textwidth, trim=10 10 10 10, clip]{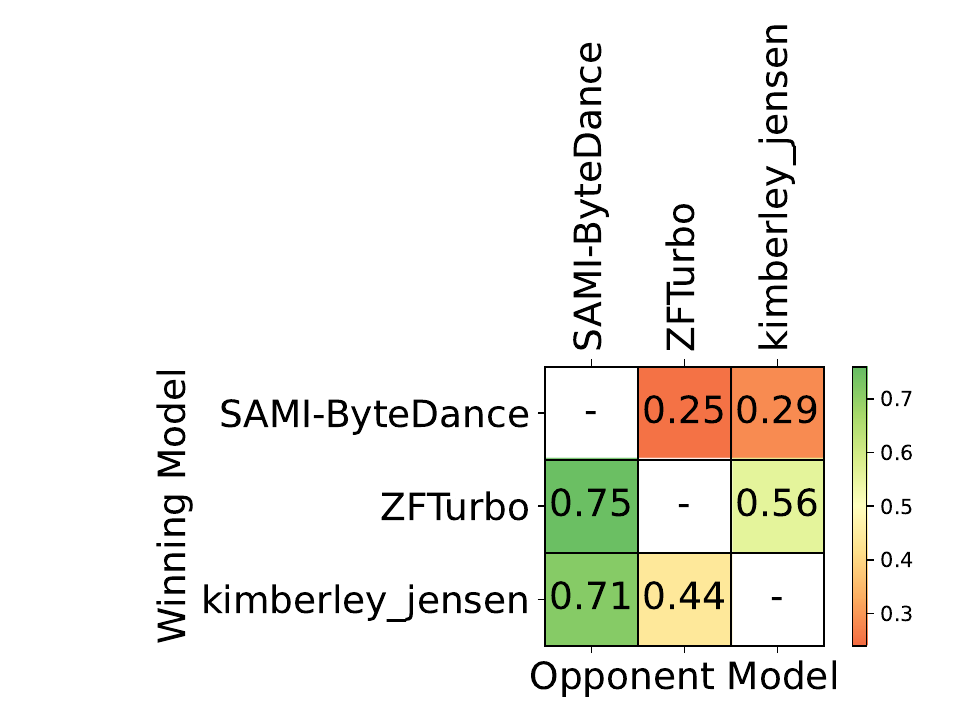}
    }
    \caption{Results of the listening test on vocal removal and extraction.}
    \label{fig:listening_test_results_vocals}
\end{figure*}

The audio segments used in the test were obtained by either extracting one of the four instruments, or by removing it.
Figures~\ref{fig:listening_test_results_bass},~\ref{fig:listening_test_results_drums},~\ref{fig:listening_test_results_other} and \ref{fig:listening_test_results_vocals} show the results on extracting and removing bass, drums, other and vocals respectively.



\subsection{TrueSkill Ratings}
In order to generate a valid global ranking of the three models, we employed the TrueSkill ranking system~\citep{TrueSkill} to summarize the results of our test.
TrueSkill generates a ranking based on a series of matches (i.e., the comparisons in our listening test) between pairs of players (i.e., the models under evaluation).
\change{Each match refines the estimation of a mean value ($0 < \mu < 50$) and a standard deviation ($\sigma$) which represent the perceived skill of the competitor (the higher the better) and a confidence on the estimation of such skill (the lower the more confident), respectively.}
\change{The procedure starts with $\mu = 25$ and $\sigma = 8.333$.}
The final ranking is shown in Table~\ref{tab:list:ratings}.
We see that the differences between the three models are very small\change{, and that the confidence is very high (due to a low $\sigma$)}.
This might be due to the models having similar performance, but also to the relatively contained size of the listening test.

Finally, TrueSkill enables us to compute the probability of a hypothetical draw between any two models.
For the match \emph{SAMI-ByteDance} vs \emph{ZFTurbo} the draw probability is 0.981, for the match \emph{ZFTurbo} vs \emph{kimberley\_jensen} it is 0.980 and for the match \emph{SAMI-ByteDance} vs \emph{kimberley\_jensen} it is 0.975.
\change{Given how high the draw probabilities between the teams are, we decided to equally split the prize pool of the listening test among the three teams.}




\begin{table}[h!]
\centering
\begin{tabular}{@{}ccccc@{}}
\toprule
 & Model & $\mu$     & $\sigma$ & SDR (Mean) \\
\midrule
1        & \emph{kimberley\_jensen}     & \textbf{24.793} & 0.779 & 9.18 \\
2        & \emph{ZFTurbo}               & 24.362 & 0.779 & 9.26 \\
3        & \emph{SAMI-ByteDance}        & 24.011 & 0.779 & \textbf{9.97} \\
\bottomrule
\end{tabular}
\caption{Final ranking obtained with TrueSkill. We used the default parameters for each player ($\mu = 25$ and $\sigma = 8.33$). \change{We report the average SDR score on leaderboard \emph{Standard} as reference.}}
\label{tab:list:ratings}
\end{table}


\subsection{Human Preference vs Objective Metric}
\begin{figure}[ht]
    \centering
    \includegraphics[width=0.45\textwidth]{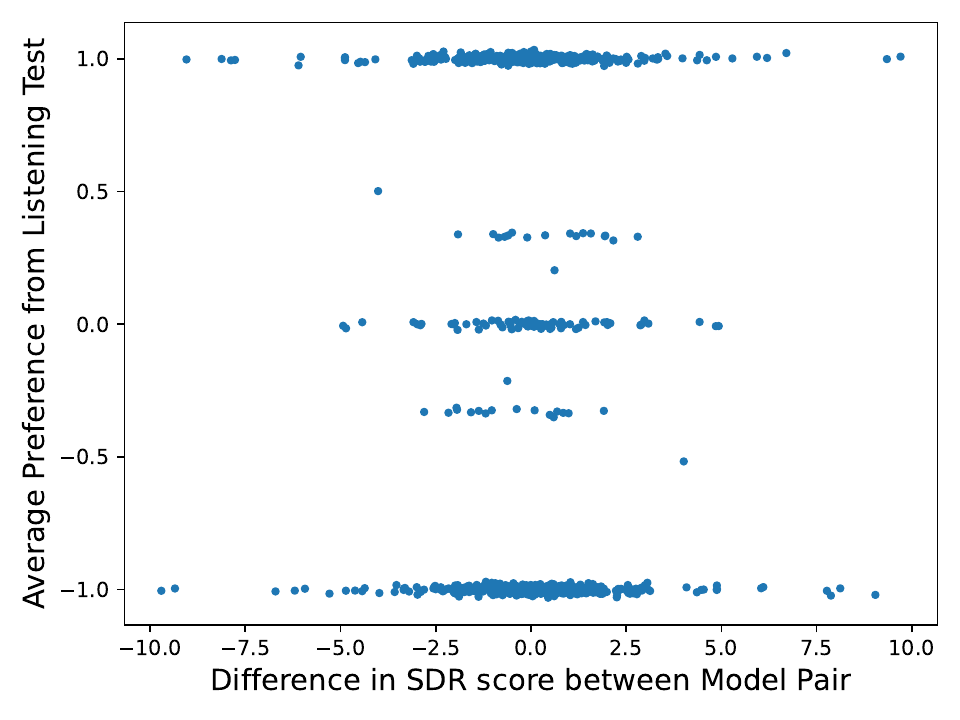}
    \caption{Correlation between the SDR scores and the results of the listening test.}
    \label{fig:sdr_listening_test_corr}
\end{figure}

Figure~\ref{fig:sdr_listening_test_corr} shows how often the human judgments agree with the objective scores based on SDR.
Each point refers to a single evaluation of the listening test.
We encode in the horizontal axis the SDR difference between the two competing models in that evaluation.
In the vertical axis we encode whether the model with higher SDR was also selected by the assessor.
As we have multiple human judgements for the same combination of model pair, song and instrument extraction or separation, we average those results.

The figure does not show a clear correlation between objective scores and the judgements made by the listeners \change{\citep[this is in line with the findings of][]{Torcoli2021ObjectiveMO}}.
The quality of all the three models participating in the listening test is high and this makes it difficult to confidently choose the best one.
Nevertheless, we find that the model by \emph{kimberley\_jensen} achieved first place in the final ranking, despite being third in the leaderboard obtained using SDR.

\section{Organizing the Challenge and Future Editions}
\label{sec:challenge_organization}
At the end of MDX'21, we stated~\citep{mitsufuji2022music} that source separation can still bring benefits to many application and research areas, and this motivated the need for future editions of the competition.
For this reason, we organized the Sound Demixing Challenge (SDX'23).
Our aim was to provide once more the familiar benchmark that participants knew already, but also expand its formulation in various directions.
At the time of MDX'21, we designed the competition so that researchers both new and experienced on this topic would be able to evaluate their models against a new test set.
We tried to keep the competition fair, by hiding the test data from the participants, and we maximized the visibility of the event by hosting it on AIcrowd\endnote{\url{https://www.aicrowd.com/challenges/music-demixing-challenge-ismir-2021}}. 

In 2023, the principles we used in organizing the challenge have not changed.
Rather, we built upon them and expanded the scope of the whole competition, to include more application areas.
Including cinematic sound separation \citep[specifically, dedicating a whole challenge track to it,][]{uhlich2023TheSoundDemixingCinematicTrack} and introducing robust \change{MSS} are two directions that we wanted to highlight as promising items for research. 
Among many reasons, this expansion has been successful also thanks to an enlarged pool of organizers and entities involved: Sony, Moises and Mitsubishi Electric Research Labs joined forces and shared efforts and resources in order to realize a bigger competition.

\subsection{Robust Music Source Separation}
Our objective of having participants develop solutions for robust training of separation models has been met only partially.
The provision of two new training sets (one for label noise and one for bleeding) has been positively received.
Given how low the availability of training data is in the research world, the participants saw this as an opportunity for more experimentation and this contributed to the success of leaderboards A and B.

On the other hand, the necessarily small size of the datasets we provided (203 songs) allowed the participants to resort to manual cleaning strategies (at least in the case of label noise): some participants spent valuable time identifying which stems were corrupted and excluded them from the dataset.
From the perspective of the challenge, this is acceptable; at the same time, though, it undermined our initial motivation for exploring such a topic, as such an activity would not scale with the amount of data and, as a consequence, would not be realistic.
Some participants reported that bleeding was clearly more difficult to tackle than label noise: we believe this to be related to the fact that \emph{SDXDB23\_Bleeding} contained exclusively problematic stems, while \emph{SDXDB23\_LabelNoise} only had approximately \SI{34}{\percent} of problematic stems.
For bleeding, no manual selection of the data was possible (i.e., the settings were more realistic): as a consequence, it was perceived as more difficult to solve.

During the course of the challenge, many participants asked us whether we would allow the usage of pretrained models available on the Internet.
We explicitly forbade the usage of such pretrained models, as that would have enabled the participants to clean the data with e.g.\ existing high quality source separation models.
This would have again undermined our initial motivation for exploring robust training of source separation models.
On top of that, although using a pretrained model is a practically viable solution for corrupted data, it would have threatened the aspects related to fairness in the competition, enabling different teams to use different tools.
For this reason, we prevented participants from potentially achieving higher scores, but ensured that everyone would use the same resources to develop their models.

\subsection{Standard Music Source Separation}
Overall, the major trend in the solutions still remains blending multiple models \citep{uhlich2017improving}.
What the participants provided were in general not individual neural networks, but many, whose outputs were combined together to create the final estimates.
We observed the same phenomenon also in MDX'21: a possibility is that, by providing trained baseline models to the participants, we implicitly encouraged this behavior.
Although in SDX'23 we provided significantly less baselines than in the previous edition, the winning models from MDX'21 were made open source after the challenge, so they represented a viable alternative to our baselines.

We noticed a larger degree of interest towards \change{the \emph{Standard}} leaderboard than towards the other two: \alter{source separation without robustness constraints} was still the most competitive playground we offered.
This allowed us to compare the evolution of the systems between the two editions of the challenge.
In MDX'21, the highest score on the final leaderboard was achieved by the team \emph{Audioshake}, with an average SDR score of \change{8.33}dB.
In SDX'23, the highest score on the final leaderboard was achieved by the team \emph{SAMI-ByteDance}, with an average SDR score of \change{9.97}dB.
In other words, the highest public score achieved on our benchmark has increased by approximately \change{1.64}dB over the course of two years.

Finally, running a full listening test on the top-performing submissions allowed us to get an alternative source of evaluation for the separation quality, besides the SDR score.
This made possible the involvement of professionals in the music industry, who represent potential users of the technology: their feedback is therefore a very important signal to take into consideration when judging the quality of a system.

\section{Conclusions}
\label{sec:conclusions}
This paper summarized the music demixing (MDX) track of the Sound Demixing Challenge 2023.
We provided a description of the challenge setup, we presented the topic of robust MSS and formalized the errors that can occur in a training dataset for source separation: label noise and bleeding.
We explained how we realized two new datasets for robust MSS: \emph{SDXDB23\_LabelNoise} and \emph{SDXDB23\_Bleeding}.
Then, we described the outcome of the challenge and reported the final results, together with a description of the winning approaches.
We detailed how the evaluation has taken place, in particular with the introduction of a listening test specifically carried out with professional figures in the music industry.
We believe that the SDX'23 challenge has given benefits to the source separation community and hope that we will continue organizing a long series of competitions in the future.

\IfFileExists{\jobname.ent}{
   \theendnotes
}{
}

\section*{Acknowledgments}
The authors would like to thank Sony Music Entertainment Japan for the creation of MDXDB21. The authors also thank the assessors who took part in the listening test.
The work of Nabarun Goswami and Tatsuya Harada was partially supported by JST Moonshot R\&D Grant Number JPMJPS2011, CREST Grant Number JPMJCR2015, and Basic Research Grant (Super AI) of the Institute for AI and Beyond of the University of Tokyo.
The work of Minseok Kim and Jun Hyung Lee was supported by the National Research Foundation of Korea (NRF) grant funded by the Korean government (MSIT) (No.\ NRF-2021R1A2C2011452).

\bibliography{bibtex/bib/IEEEabrv,bibtex/bib/paper}

\end{document}